\documentclass[sigconf]{acmart}

\usepackage{placeins}
\usepackage{enumitem}       %
\usepackage{float}          %

\AtBeginDocument{%
  }

\copyrightyear{2026}
\acmYear{2026}
\setcopyright{cc}
\setcctype{by}
\acmConference[CHI '26]{Proceedings of the 2026 CHI Conference on Human Factors in Computing Systems}{April 13--17, 2026}{Barcelona, Spain}
\acmBooktitle{Proceedings of the 2026 CHI Conference on Human Factors in Computing Systems (CHI '26), April 13--17, 2026, Barcelona, Spain}
\acmPrice{}
\acmDOI{10.1145/3772318.3790893}
\acmISBN{979-8-4007-2278-3/2026/04}

\acmSubmissionID{8162}

\usepackage{multirow}
\usepackage{indentfirst}

\begin{document}

\title{Where Digital Meets Place: Deriving Strategies for Curating Mixed Reality Exhibitions in Public Spaces}

\author{Yawei Zhao}
  \orcid{0000-0001-9298-0955}
\affiliation{%
  \institution{The Hong Kong University of Science and Technology (Guangzhou)}
  \city{Guangzhou}
  \country{China}
}
\email{yzhao099@connect.hkust-gz.edu.cn}
 
\author{Jiaxin Liang}
\orcid{0000-0003-1062-1970}
\affiliation{%
  \institution{The Hong Kong University of Science and Technology}
  \city{Hong Kong}
  \country{China}
}
\email{jliangbh@connect.ust.hk}

\author{Hao Li}
  \orcid{0009-0007-2434-4325}
\affiliation{%
  \institution{The Hong Kong University of Science and Technology (Guangzhou)}
  \city{Guangzhou}
  \country{China}
}
\email{hli307@connect.hkust-gz.edu.cn}

\author{Pan Hui}
\orcid{0000-0001-6026-1083}
\affiliation{%
  \institution{The Hong Kong University of Science and Technology (Guangzhou)}
  \city{Guangzhou}
  \country{China}
}
\affiliation{%
  \institution{The Hong Kong University of Science and Technology}
  \city{Hong Kong}
  \country{China}
}
\email{panhui@ust.hk}
\renewcommand{\shortauthors}{Zhao et al.}

\begin{abstract}
 Mixed Reality (MR) technologies are increasingly being used to enrich exhibitions and public spaces by blending digital content with the physical environment in real time. However, little is known about curatorial strategies for embedding MR exhibitions into public spaces or promoting audience experiences. To explore this, we designed and curated a campus-based MR art exhibition, using contextualism as the fundamental concept. We conducted an interdisciplinary expert focus group alongside exhibition viewing to identify opportunities, challenges, and design strategies from multiple perspectives. In parallel, we conducted user studies with general audiences to examine how curatorial strategies foster experiential qualities. Our findings reveal insights from both experts and general users along with strategies in curating MR exhibitions and highlight the foundational role of contextualism in curating MR art exhibitions in urban public spaces.
\end{abstract}




\begin{teaserfigure}
  \includegraphics[width=\textwidth]{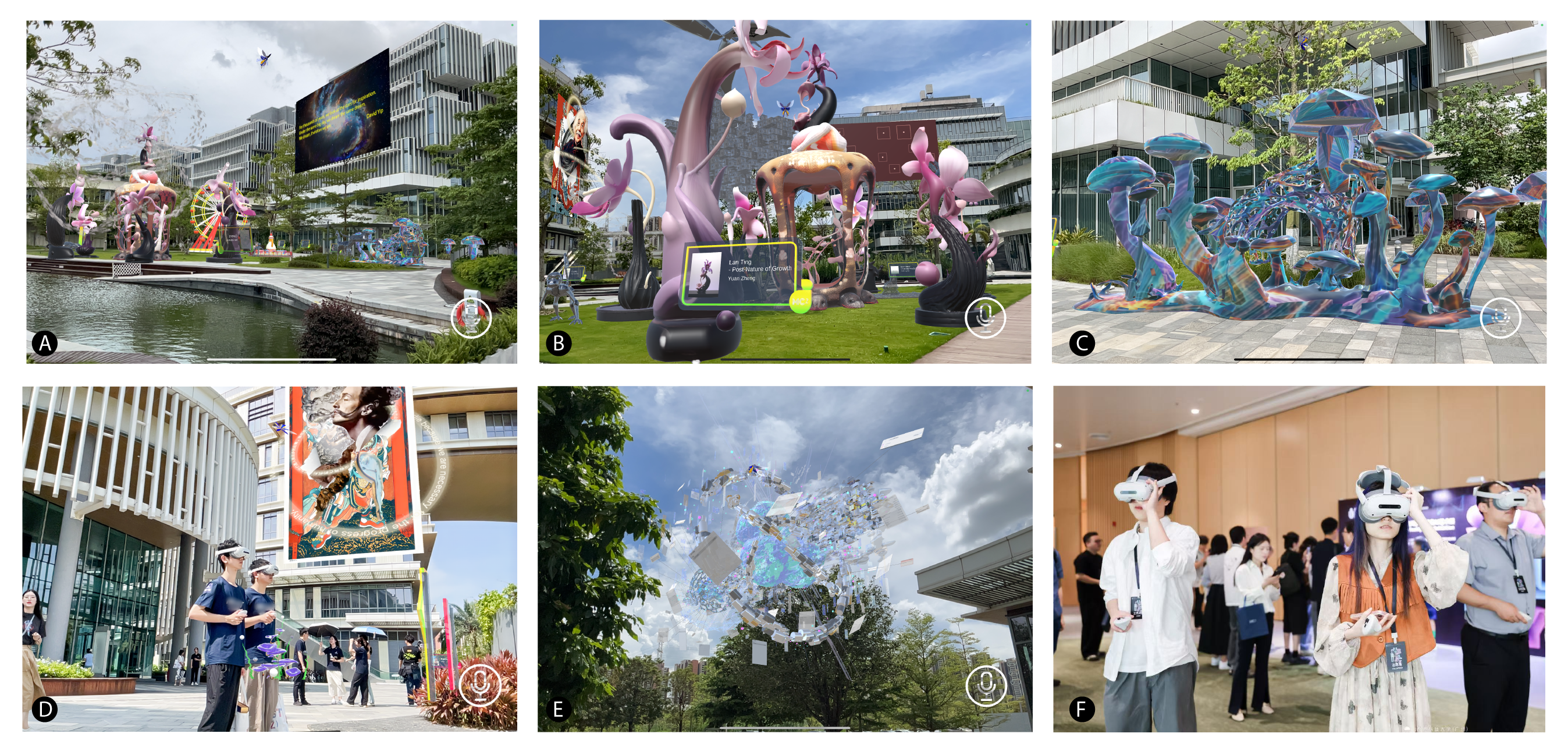}
  \caption{Scenes from the campus-based mixed-reality art exhibition. (A–E) Headset views of digital artworks anchored in different campus locations, blending with the physical environment. (F) On-site visitors using head-mounted displays to experience the exhibition.}
  \Description{A teaser figure containing six thumbnails arranged sequentially from upper left to bottom right. Thumbnails A–E show mixed reality artworks situated in real-world public spaces, while F depicts visitors wearing head-mounted devices as they view the exhibition.}
  \label{fig:teaser}
\end{teaserfigure}

\maketitle 

\section{Introduction}
\label{Sec: Intro}
Mixed Reality (MR) technologies are increasingly used to enrich public spaces by blending digital content with the physical environment in real time \cite{benford2011performing, rokhsaritalemi2020review}. Within Human–Computer Interaction (HCI), MR has emerged as a promising medium for creating immersive and interactive experiences that extend beyond traditional screens and devices \cite{huang2016survey,papadopoulos2021interactions}. Cultural institutions, such as museums and galleries have begun experimenting with MR to enhance guidance, provide interpretive functions, enable new forms of audience engagement, and make art more accessible to diverse audiences \cite{hammady2021framework,xiao2024application}. Prior studies show that MR can increase engagement and support learning by fostering immersion, presence, and contextualized storytelling \cite{hammady2020ambient}. These developments highlight the potential of MR for spatial storytelling, where the environment itself becomes a narrative canvas that fosters immersion and participation \cite{mcquire2008mediacity,Smed_2021}. At the same time, they signal a growing interest in how MR can transform not only the way audiences encounter artworks but also the curatorial practices that shape those encounters \cite{yeom2021digital, Murchu2016}.

Prior work on MR-based exhibition practices has demonstrated technical feasibility, introduced novel interaction techniques, and explored initial user experiences in museums and galleries \cite{Chng_2020_AcceptanceExperienceExpectationsCultural,Jung_2016_AugmentedRealityVisitorExperiences}. However, most MR installations have been developed for single, localized sites—often within museums or heritage contexts where the physical setting already carries cultural meaning or narrative significance \cite{Shin_2023_LinkingTrajectoryNarrativeIntent}.
And the existing guidelines, originating from this context, are effective for interpretive storytelling \cite{Millard_2020_CreatingLocativeCulturalStorytelling,Shin_2018_DigitalHeritageStorytellingTool, Rizvic_2014_CulturalHeritageApplications}, but offer limited guidance for art exhibitions that curate born-digital works across urban environments not anchored by prior location-based meaning. 

Therefore, digital art exhibitions, especially those situated in open public spaces, require a distinct curatorial logic. Insights from large place-based art festivals and exhibitions \cite{kaye2013site,Lynch_2022_festivalmakingartisticproduction} show that orchestrating spatial flow, thematic cohesion, and technological consistency is essential for supporting visitors’ immersion in the artworks. For emerging MR art exhibitions, these considerations become equally important, as the medium introduces new spatial and experiential conditions that must be curated with similar care \cite{Jiang_2025_user-centreddigitalartdesign, Murchu2016}. This calls for rethinking not only the spatial deployment of MR content but also the interdisciplinary processes through which curators, designers, and technologists collaborate to shape engagement across diverse environments \cite{Damala_2012_RealityA2Rmuseumvisit, Lee_2022_teamLab}. To address this gap, we pose the following research questions:

\begin{itemize}

    \item \textbf{RQ1:} What opportunities and challenges arise when curating an MR art exhibition within existing public space environments?
    \item \textbf{RQ2:} From interdisciplinary professional perspectives (curators, artists, urban designers, and MR technologists), what design strategies can address the challenges of curating MR exhibitions in existing physical environments?
    \item \textbf{RQ3:} What strategies can effectively foster user experiences such as engagement, embodiment, and immersion for the general audience in MR exhibitions?
\end{itemize}

To answer these questions, we curated and implemented a campus-based MR art exhibition, adopting contextualism as our curatorial and design foundation, which views place as integral to the MR experience \cite{Lynch_1960,Elshater_2025_FormationTextualRepresentation}. The exhibition was conceived as an opportunity to investigate the curatorial, spatial, and experiential dimensions of MR in a real-world public space. We convened an interdisciplinary expert focus group—bringing together curators, artists, architects/urban designers, and MR technologists—to reflect on opportunities and challenges. In parallel, we conducted a user study with participants from the general audience to gather feedback on experiential aspects. Together, these studies enabled us to examine MR exhibition design as both a curatorial process and a lived audience experience, combining contextualism-guided MR exhibition design flow with insights from both experts and the general audience. This multi-stage approach allowed us to analyze MR curation through (1) curatorial decisions made in situ, (2) expert insights across professional perspectives, and (3) general audience reflections during the exhibition.

Through this study, we contribute to the field of HCI with (1) a set of curatorial strategies for embedding MR exhibitions in public spaces, offering guidelines for integrating digital layers with public space contexts; (2) a reflection on MR exhibition design, illustrating how contextualist concepts can structure MR art exhibition curation; and (3) insights and evaluations from distinct audience groups, providing practical recommendations for curators, designers, and technologists to enhance experiential quality of MR exhibitions.

\section{Background and Related Works} \label{sec: review}

Mixed Reality (MR)—often encompassing Augmented Reality (AR) and related technologies—refers to the blending of digital content with the physical environment in real time. Within HCI, MR has gained momentum as a means to enrich exhibitions and physical spaces, creating phygital experiences that merge tangible surroundings with virtual information \cite{Maddali_2023_whenReconstructingMeaningfulSpaces, devrio2025reel}. In recent years, rapid advances in MR hardware (e.g., mobile AR, headsets) and software have enabled museums, galleries, and public venues to experiment with new forms of visitor engagement. Foundational work on the Reality–Virtuality Continuum \cite{milgram1994taxonomy} laid the conceptual basis for MR, but more recent studies have examined its practical impact on visitor experience, learning, and social interaction in exhibition contexts \cite{Bareisyte_2024_realitysystematicscopingreview}.

This section reviews work across three themes: (1) place-based art festivals and exhibitions, which offer curatorial precedents for staging art in large-scale environments; (2) MR-based exhibition practices, spanning applications from small, object-centered installations to distributed experiences in public space; and (3) bringing contextualism as a theoretical guidance for curating MR art exhibitions. Together, these themes frame our research questions and design foundations for this study.

\subsection{Place-Based Art Festivals and Exhibitions }\label{sec:art festival}
Place-based (and often site-specific) art festivals—such as rural and island triennials—offer a mature curatorial model for staging art in real environments, such as the Echigo-Tsumari Art Triennale (ETAT) \cite{echigo} and Setouchi Triennale \cite{setouchi}. Rather than treating the gallery as a neutral container, these programs distribute works across landscapes, streetscapes, and community infrastructures, asking audiences to move, navigate, and make sense of artworks in relation to the character of place \cite{Lynch_2022_festivalmakingartisticproduction, Cai_2020_TourismSustainableTourismEconomics}. In art-historical terms, this aligns with site-specificity (artworks conceived for, and inseparable from, a particular site) and with contemporary place-based cultural strategies that braid artistic intent with local narratives, mobility, and stewardship \cite{kwon2004one,kaye2013site,bishop2023artificial}.
Place-based festivals typically curate at multiple spatial scales: the territory (routes, hubs, and wayfinding), the site (micro-settings, materials, acoustics, lighting), and the work (form, interaction, and durational rhythms) \cite{Lynch_2022_festivalmakingartisticproduction}. Wayfinding maps, checkpoints, and narrative clusters guide attention without over-determining paths, balancing open wandering with essential “must-see” anchors. Information is layered—on-site signage, guidebooks, docents—so that interpretation unfolds in situ, with the environment acting as both frame and medium \cite{saayman2012finding}.

These festivals also demonstrate how art can activate large sites and regions, offering precedents directly relevant to MR \cite{Widjaja_2025_RealityMixedRealityExhibition}. As exhibitions increasingly adopt MR, digital layers can extend what physical media alone cannot—lifting material constraints to deepen immersion and situate attention. Our campus-based MR exhibition builds on this lineage to examine MR’s capacity for large-site storytelling and its practical feasibility in public-facing contexts. The next section reviews MR-based exhibition practices—drawing out recognized location-based strengths alongside recurrent limitations.

\subsection{MR-Based Exhibition Practices} \label{sec: review MR}
Over the past decade, mixed reality (MR)—including augmented reality (AR), virtual reality (VR), projection mapping, and sensor-based interaction—has been widely adopted across museums, galleries, and cultural-heritage contexts. Prior studies consistently show that MR can heighten immersion and presence, increase visitor engagement, and support learning through contextual visualizations and interactive storytelling \cite{Chng_2020_AcceptanceExperienceExpectationsCultural,Jung_2016_AugmentedRealityVisitorExperiences,Damala_2012_RealityA2Rmuseumvisit}.  In practice, in-gallery AR/VR interactions continue to evolve: for example, optimizing pointing and selection for head-mounted displays to enable more natural browsing and manipulation in situated exhibitions \cite{abe2025understanding}; re-constructing historical and media contexts in immersive environments to strengthen links between artifacts and narrative \cite{Alliata_2024_CinemaImmersiveEnvironment,Hu_2024_AugmentedRealityEnvironmentInteractive}. 

Despite these advances, persistent limitations remain. Issues of usability and robustness—including cybersickness, arm fatigue, interaction friction, and tracking or registration errors—continue to disrupt experiential continuity  \cite{Ratcliffe_2021_RemoteResearchSurveyDrawbacks,Luo_2025_ExploringInteractionTechniquesSpatial}. Some scholars also identify safety concerns associated with heavy reliance on head-mounted devices \cite{Sharma_2017_SpaceMakePlaceSpatial}. Additionally, information layering and spatial coherence pose significant challenges. As Millard argues \cite{Millard_2020_CreatingLocativeCulturalStorytelling}, achieving \textit{loco-narrative harmony}—a deliberate alignment between the physical site and the narrative presentation—requires careful spatial and dramaturgical design so that virtual elements enhance rather than fragment visitors’ attentional balance  \cite{Bryan-Kinns_2023_MediaArtsTechnology,Alliata_2024_CinemaImmersiveEnvironment,Luo_2025_ExploringInteractionTechniquesSpatial}. 

Beyond indoor exhibitions, MR has increasingly moved into plazas, streetscapes, and natural sites, aligning with trajectories in media architecture \cite{Houben_2017_MediaArchitecture} that treat buildings and city surfaces as interactive, sensor-driven canvases \cite{Muller_2010_interactivepublicdisplays}. Outdoor MR installations have been shown to activate public spaces, catalyze social gathering, and foreground place-based qualities such as affordances, flow, and ambiance \cite{Kukka_2017_artexhibitionsinteractivepublicb, Papageorgopoulou_2021_installationbuildingenhancingcitizens}. Recent work links spatial augmentation to experiential shifts at the urban scale, from landscape–perception coupling to well-being in digital art exhibitions \cite{Fang_2024_assessmentstreetscapeperspectivelandscape-perception, Jiang_2025_user-centreddigitalartdesign, Xia_2024_psychologicalwell-beingChineseGeneration}. However, most of these MR installations are designed for a single, localized site. Only a small number of studies address distributed or multi-site MR experiences, such as Fenu’s \textit{Svevo Tour} \cite{Fenu_2018_experimentationaugmentedrealityapplication} or Shin’s focus group study guidelines \cite{Shin_2018_DigitalHeritageStorytellingTool}.
These works primarily focus on narrative hotspots and heritage meaning-making rather than curatorial integration for MR-based art exhibitions.

On the other hand, research has begun to theorize how virtual content and physical context co-construct spatial experience. Shin et al. \cite{Shin_2024_NarrativeSpacesVirtual-RealConnections} introduce augmented narrative spaces, showing that narrative perception in MR is shaped not only by digital content but also by spatial affordances, environmental constraints, and users’ embodied trajectories \cite{Benford_2009_coherentjourneysuserexperiences}. Some case-based MR studies underscore that curating MR in physical space is inherently interdisciplinary, requiring curators, artists, architects/urban designers, and HCI technologists to work as one team \cite{Numan_2025_OutdoorAugmentedRealityExperiences,Lee_2022_teamLab,Gao_2023_VisitingExperienceUsingAR}. Heritage and gallery projects highlight the need for integrated pipelines—covering content modeling, interaction design, spatial integration, and deployment—while contemporary outdoor AR projects often rely on collaboration-heavy authoring processes \cite{Bozzelli_2019_frameworkuser-centricinteractiveexperience, Franz_2019_Table, Ppali_2025_ImmersiveExperiencesArt-MakingPeoplea}.

Despite growing interest in MR across both indoor and outdoor environments, curatorial methods for MR art exhibitions remain underdeveloped. Existing guidelines largely stem from museum and heritage AR, where digital overlays reinforce pre-existing narratives embedded in specific locations \cite{Shin_2024_NarrativeSpacesVirtual-RealConnections,Bozzelli_2019_frameworkuser-centricinteractiveexperience}. These logics differ from art exhibitions, which must curate born-digital artworks across sites that do not inherently possess fixed historical or interpretive meaning \cite{Kreps_2013_PerspectivesMuseumsCurationHeritage}. 

Prior work demonstrates that MR can reshape exhibition experiences across different scales—from single-gallery installations to distributed sites in large urban environments. 
However, empirical work on how to curate and coordinate MR artworks across multiple real-world sites, particularly in non-museum art environments, remains limited. In our MR Art Exhibition, we document a complete pipeline that includes site and context analysis, cross-disciplinary collaboration, technical integration, and on-site deployment. Building on prior research on MR exhibitions, our goal is to extend this foundation by developing curatorial strategies suited to large-scale, real-world MR art environments and aligned with the capabilities and constraints of emerging MR technology.

\subsection{Bringing Contextualism to MR Exhibition } \label{sec: review context}

In this study, we bring contextualism as a conceptual framework to guide the design of our MR art exhibition in urban public space. And we aim to develop it further in the case of curating urban public space MR exhibition. Contextualism emphasizes that meaning is inherently situated—shaped by the physical, social, and cultural environment in which an experience unfolds. This perspective has been widely discussed across HCI, art, urban design, and curatorial studies. Drawing on these interdisciplinary foundations, we position contextualism as central to how digital content, site, and visitor engagement are integrated in our curatorial approach. In the following section, we unpack how contextualism has been interpreted across disciplines. This helps clarify why we adopt contextualism as the guiding concept for our MR exhibition.

In HCI, contextualism is reflected in theories such as situated action \cite{suchman1987} and situated cognition \cite{brown1989situated}, both of which emphasize that behavior and meaning emerge from real-time interactions within specific environments. These views highlight the importance of designing MR experiences that are responsive to place and context. In our MR exhibition, these theories informed our understanding of how site-specific conditions shape user experience, reinforcing our adoption of contextualism as a guiding framework.

Similar ideas appear in architecture and the arts. The concept of genius loci—the “spirit of place”—in architecture emphasizes that design should respond to the unique character of a site \cite{norberg1980genius, vecco2020genius}. In contemporary art, site-specificity describes practices where artworks are conceived in dialogue with a location’s historical and spatial conditions \cite{kaye2013site, Bedford_2016_}. Both perspectives position the physical setting as an active agent in shaping meaning and experience, resonating with contextualist thinking in their emphasis on place-aware creation.

In summary, contextualist thinking infuses modern MR exhibition design with a focus on place-specific meaning and situated experience. It bridges theory and practice: from HCI’s emphasis on situated action \cite{suchman1987} to architecture’s respect for genius loci \cite{norberg1980genius}and art’s commitment to site-specificity, these ideas collectively encourage MR creators to design with context in mind. contextualist thinking infuses modern MR exhibition design with a focus on place-specific meaning and situated experience. By translating contextualist thinking into curatorial practice, this study contributes to the development of MR design strategies that are both technologically robust and experientially grounded in place.

\section{Methodological Approach}

\begin{figure*}[h]
  \centering
  \includegraphics[width=0.85\textwidth]{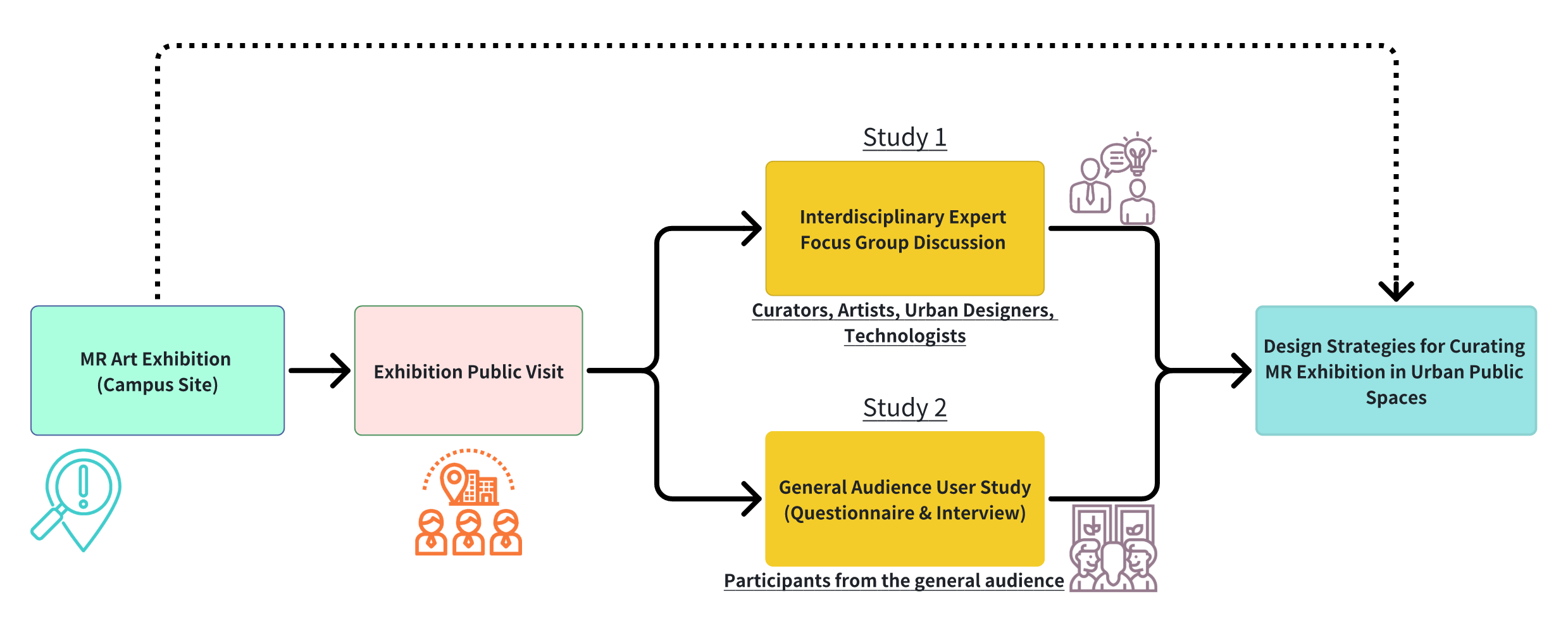}
  \caption{Overview of the Mixed-Methods Framework.}
  \Description{A study flow diagram illustrating the process: spatial analysis and exhibition curation lead to the exhibition’s public viewing, which in turn informed Study 1 (Expert Focus Group) and Study 2 (User Study). Together with the site analysis and exhibition curation, these two studies converged to generate novel design implications for curating MR exhibitions in urban public spaces.}
  \label{fig:stud flow}
\end{figure*}

This research employed a design-led mixed-methods framework consisting of multiple stages (see Figure~\ref{fig:stud flow}). First, an MR art exhibition was curated and implemented in a campus public space. Guided by contextualism as a conceptual foundation \cite{Elshater_2025_FormationTextualRepresentation}, site-responsive curatorial strategies were developed to embed MR interventions into the site’s spatial narratives. The exhibition was presented to the public through guided tours, establishing an in-situ context and empirical basis for subsequent evaluation.

Building on this exhibition, two empirical studies were conducted during the exhibition period: an interdisciplinary expert focus group study and a user study with members of the exhibition’s general audience. The expert study took the form of a focus group with experts from different professional backgrounds, while the user study involved a questionnaire and semi-structured interviews with participants recruited from general visitors who had experienced the exhibition. Details of the participants, procedures, and analyses for both studies are provided in Section~\ref{Section5_Study1_Method} and Section~\ref{Section6_Study2_Method}.

Within this methodological framework, the exhibition design and the two empirical studies play complementary roles in addressing the three research questions introduced in Section~\ref{Sec: Intro}. The exhibition design and curatorial process provide design-side insights into the opportunities and challenges of embedding MR interventions in an existing physical environment (\textbf{RQ1}). Study~1, the interdisciplinary expert focus group study, further elaborates these opportunities and challenges and develops curatorial design strategies from professional viewpoints (\textbf{RQ1} and \textbf{RQ2}). Study~2, the user study with members of the exhibition’s general audience, examines how the implemented strategies shape visitors’ experiences such as engagement, immersion, and sese of presence, thereby contributing to \textbf{RQ3}.

Empirical material was collected and analyzed across components in complementary ways. Expert discussions and general user interviews were thematically analyzed to identify recurring opportunities, challenges, and insight patterns. The responses of the user study questionnaire were quantitatively analyzed with descriptive statistics and Cronbach’s $\alpha$ reliability analysis, providing an overview of the’ evaluations that complements the qualitative insights.

Taken together, these stages enabled MR exhibition design to be examined from three complementary perspectives: as reflective design decisions \cite{ghajargar2021synthesis}, as assessed by professionals \cite{roedl2013design}, and as experienced by the general audience \cite{mcveigh2019shaping}. All procedures involving human participants were reviewed and approved by the university's Research Ethics Committee. All participants provided informed consent before taking part in the study. Data collection was conducted anonymously, with no personally identifiable information recorded or reported.

\section{The MR Exhibition Design}
\begin{figure*}[t]
    \centering
    \includegraphics[width=\linewidth]{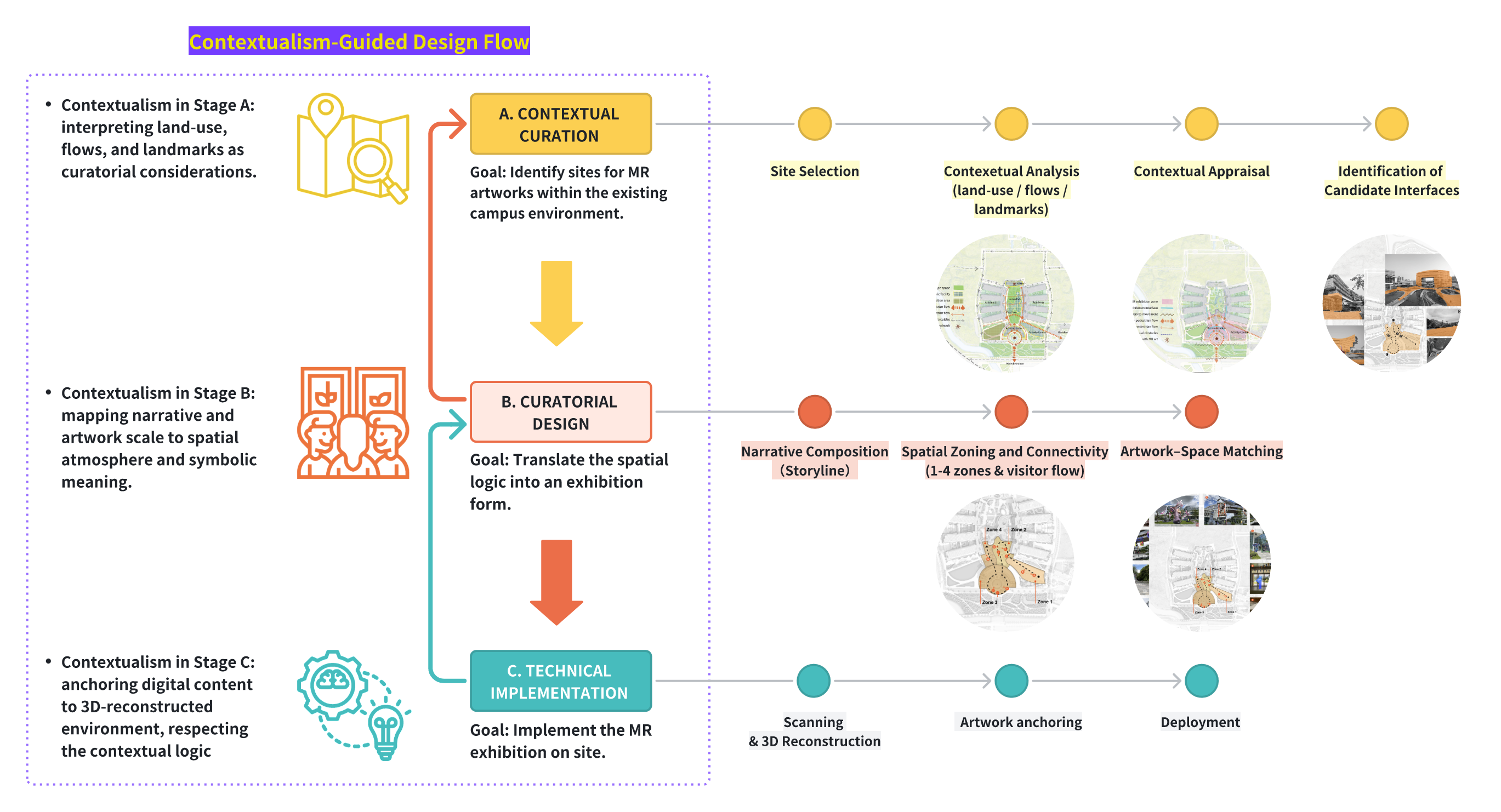}
    \caption{Contextualism-guided MR exhibition design flow. Contextualism operates as the overarching principle that links contextual curation (A), curatorial design (B), and technical implementation (C).}
    \label{fig:designflow}
\end{figure*}

This section presents the design of the MR exhibition as a contextualism-guided process (Figure~\ref{fig:designflow}). The design flow was structured into three stages: A. \textit{Contextual Curation}, B. \textit{Curatorial Design}, and C. \textit{Technical Implementation}. Stage A interpreted land uses, pedestrian flows, and campus landmarks to identify suitable sites for MR intervention. Stage B translated this spatial logic into an exhibition form by mapping narrative themes and artwork scales onto the selected spaces. Stage C implemented the MR exhibition on site by deploying the MR system so that digital content aligned with the physical environment. The following subsections elaborate on each stage in turn, detailing how contextual curation, curatorial design, and technical implementation jointly shaped the final exhibition layout and audience journey.

\subsection{Contextual Curation}
Contextual curation applies contextualism as a curatorial method by using the existing spatial organisation and everyday practices of the campus as the primary basis for deciding where MR interventions should be located in relation to campus life. To operationalise this approach, contextual curation combined \textit{Site Selection}, \textit{Contextual Analysis}, \textit{Contextual Appraisal}, and the subsequent \textit{Identification of Candidate Areas} into a structured workflow in which the first three stages delineated suitable zones within the campus core, and the final stage refined these zones into concrete surfaces and spatial interfaces for hosting MR artworks.


\subsubsection{Site Selection.}
Site selection focused on locating the exhibition in areas that were publicly accessible, embedded in everyday campus activities, and capable of supporting sustained pedestrian flow. Guided by a functional zoning analysis, residential quarters, sports facilities, and areas under construction or pending development were excluded. This process led to the identification of the campus core as the primary setting for the MR exhibition, as it concentrates academic, administrative, and student-life functions while remaining open to the wider public.

\subsubsection{Contextual Analysis.}
\begin{figure}[h]
  \centering
  \includegraphics[width=\linewidth]{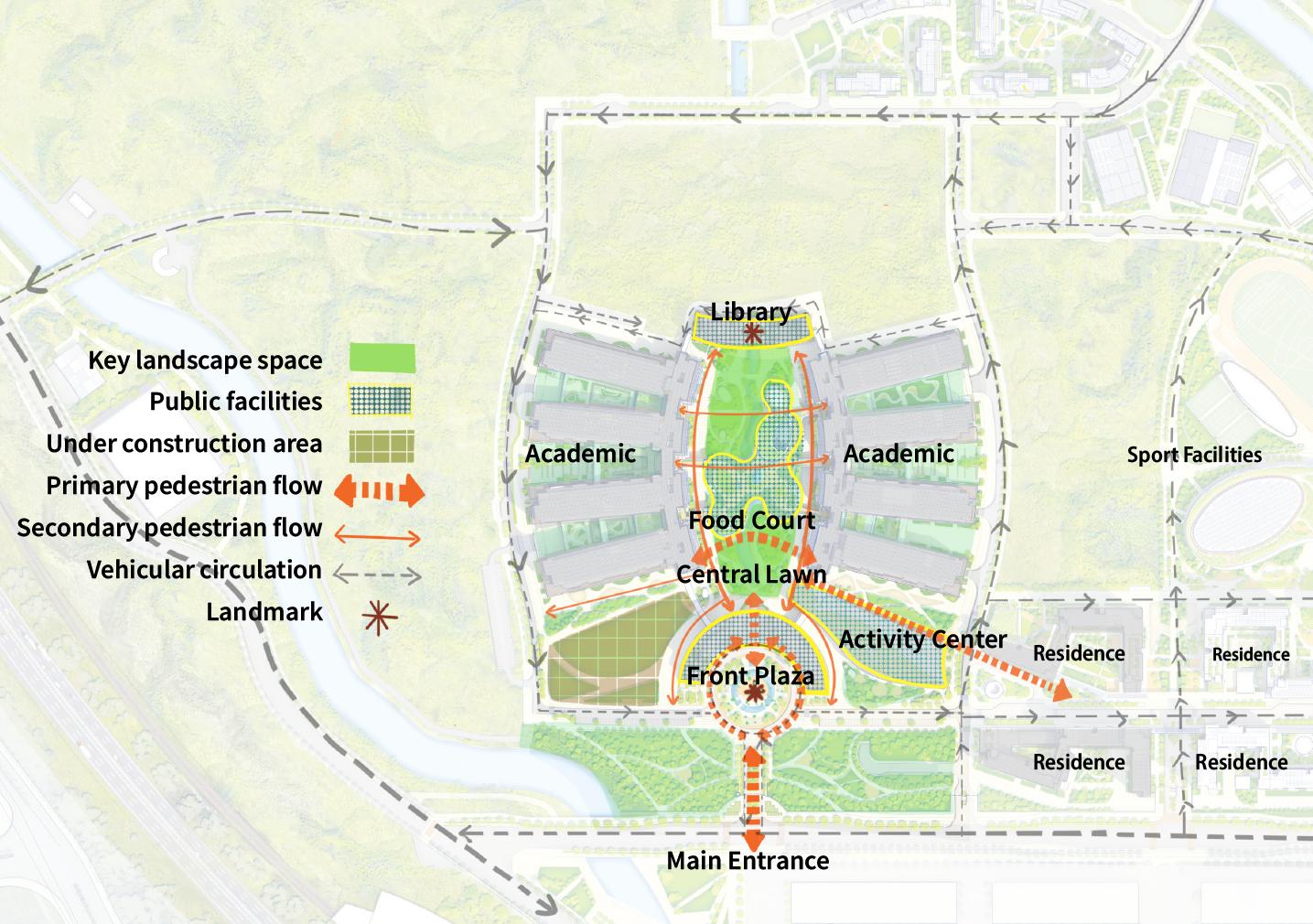}
  \caption{Contextual analysis of the campus, illustrating how functional zones (land uses), circulation flows, and key landmarks contribute to the overall legibility of the site.}
  \Description{A design analysis diagram illustrating the campus’s functional zones (land use), circulation flows, and functional zones.}
  \label{fig:ContextualAnalysis}
\end{figure}

The contextual analysis organised the campus core into patterns of spatial arrangement and use, so that subsequent curatorial decisions could be grounded in a structured understanding of the site. Following urban design work that emphasises contextual sensitivity and spatial legibility \cite{carmona2014place}, the analysis focused on three complementary dimensions: land uses, circulation flows, and key landmarks (as shown in Figure~\ref{fig:ContextualAnalysis}) \cite{Lynch_1960, Lynch_2022_festivalmakingartisticproduction}. Considered together, these dimensions revealed a distinct functional structure within the campus core: corridors that primarily serve everyday circulation, open spaces that act as multi-purpose activity nodes, and entrance areas that function as symbolic gateways. This means that contextual analysis operates as a methodological foundation for the subsequent appraisal stage by characterizing the spatial roles of different parts of the campus and treating them as potential settings for MR exhibition spaces.

\subsubsection{Contextual Appraisal.}
Contextual appraisal was built on the contextual analysis by evaluating which parts of the campus core could feasibly and appropriately be transformed into MR exhibition settings. Drawing on contextualist urban design literature that links patterns of use, circulation, and symbolic form to how people appropriate and experience places \cite{carmona2014place,Lynch_1960}, the appraisal considered three main criteria: (i) whether an area supports public-facing and diverse activities, so that MR artworks can attach to existing campus social life; (ii) its exposure to everyday circulation, so that MR content can be encountered as part of routine movement rather than requiring a special trip; and (iii) its symbolic prominence within the campus landscape, so that MR interventions can engage with and reflect the representational identity of the university.

\begin{figure}[h]
  \centering
  \includegraphics[width=\linewidth]{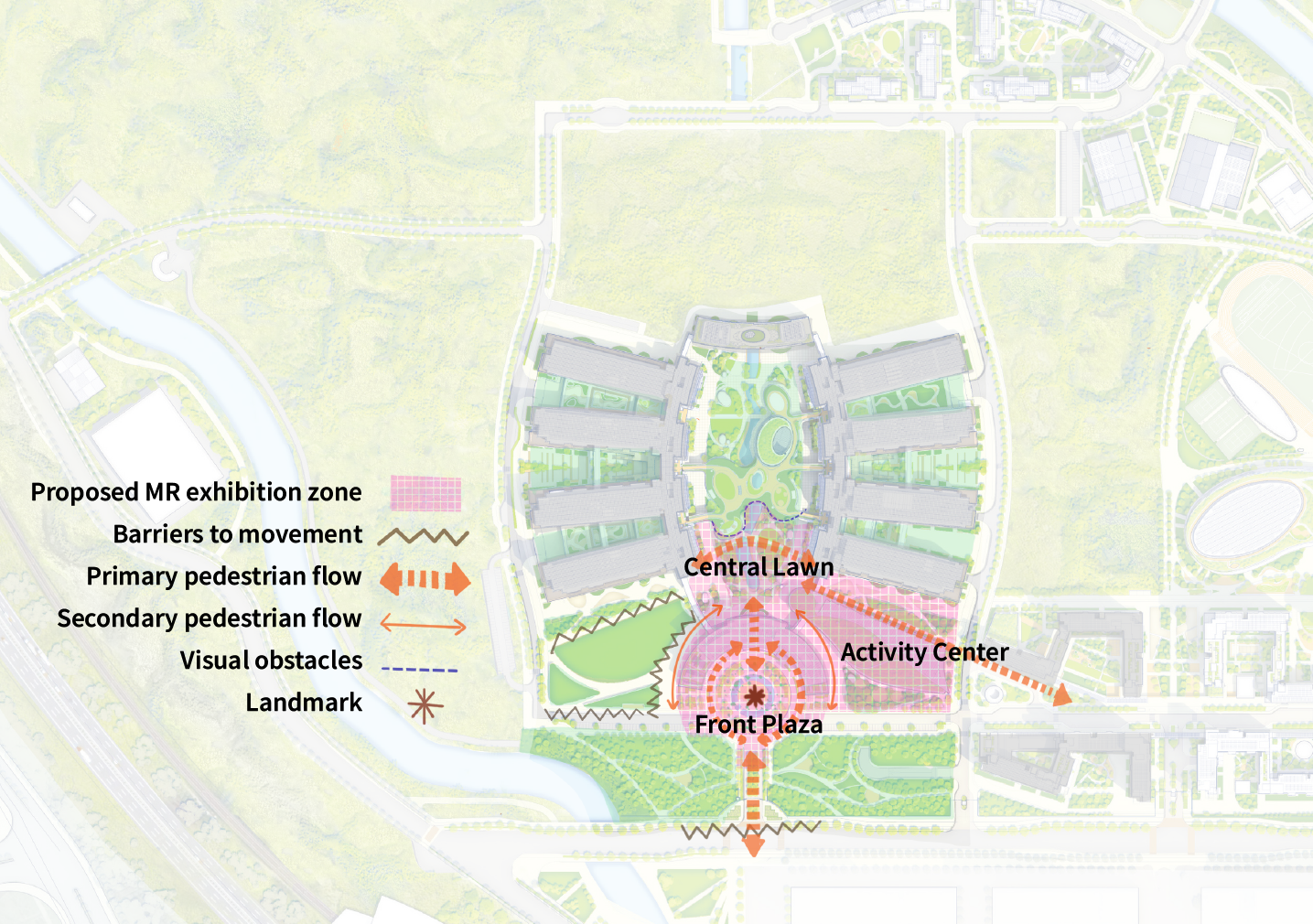}
  \caption{Contextual appraisal for the proposed MR exhibition, identifying opportunities and limitations for embedding MR experiences into the site by mapping flows, barriers, and obstacles.}
  \Description{A contextual appraisal diagram mapping flows, barriers, and obstacles across the site.}
  \label{fig:ContextualAppraisal}
\end{figure}

Through this appraisal process, three primary areas were therefore selected as potential exhibition sites (Figure~\ref{fig:ContextualAppraisal}): the Student Activity Center, the Central Lawn quadrangle, and the Front Plaza. Conceptually, these areas instantiate three spatial roles within the campus core—an indoor activity hub, an open movement field, and a symbolic gateway—which provide the basis for the subsequent identification of concrete surfaces and spatial interfaces capable of hosting MR artworks.

\subsubsection{Identification of Candidate Interfaces.}
\begin{figure}[h]
  \centering
  \includegraphics[width=\linewidth]{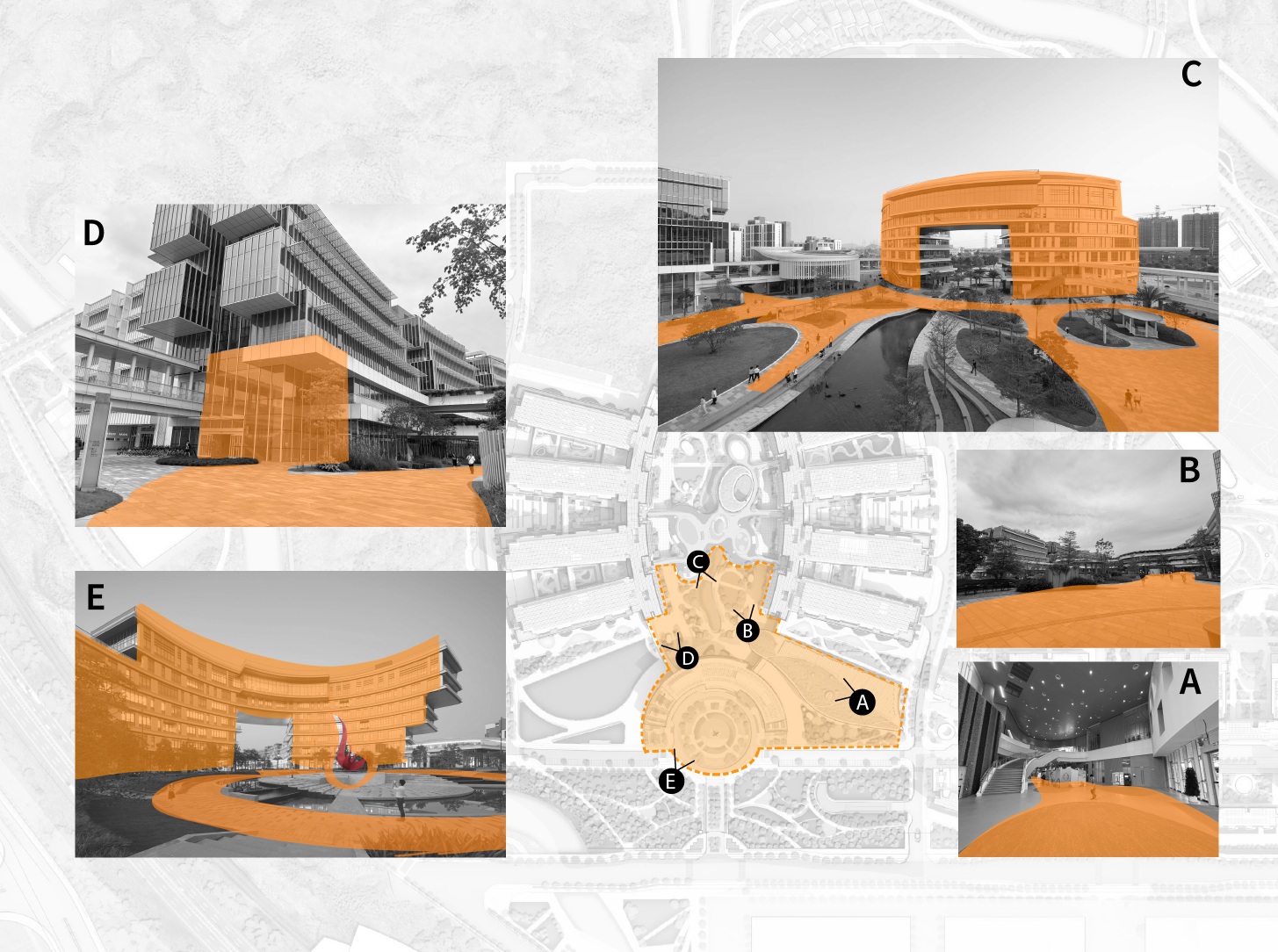}
  \caption{Candidate interfaces. Highlighted zones indicate the surfaces and volumes where MR content can be accommodated (A-E). A shows a perspective from the indoor activity center; B and D present human-level views directed toward the central lawn ground plane; C shows a bird’s-eye view of the central lawn area; and E shows a perspective toward the front plaza.}
  \Description{A campus map with thumbnails showing different spatial perspectives. Potential interfaces for embedding digital artworks are highlighted in color.}
  \label{fig:Candidateinterface}
\end{figure}

Building on the previously appraised areas, this stage focused on how specific surfaces and volumes within each area could accommodate MR content. Moving from a plan-based analysis of the campus to a three-dimensional reading, potential interfaces were identified in terms of their capacity to host virtual overlays and their qualities from a visitor’s point of view—namely visibility \cite{yang2007viewsphere} (vertical sightlines) and accessibility \cite{koenig1980indicators} (Figure~\ref{fig:Candidateinterface}). These interfaces included interior floors and high-ceilinged spaces in the Student Activity Center, ground planes and overhead volumes along the Central Lawn, and focal regions around the Front Plaza sculpture and façade. This step translated the area-level roles of indoor activity hub, open movement field, and symbolic gateway into situated interfaces that subsequently structured how individual MR works were positioned in the curatorial design.

\subsection{Curatorial Design}

The goal of curatorial design stage is to organize how MR works are distributed with respect to the identified campus interfaces. In this stage, the areas appraised in the previous steps are further refined into specific exhibition zones and connected through a designed movement flow, preparing them for artwork–space matching. 

There were three steps in this stage (Figure~\ref{fig:designflow}). First, the narrative composition within the collected artworks was identified, clustered into thematic groups, and connected into a loose storyline. Then, at the same time, the appraised areas were redefined into four zones and integrated with the exhibition flow, determined by functional zoning and connectivity. As a result, the thematic groups were aligned with exhibition zones. Third, artwork–space matching assigned individual MR works to specific interfaces within each zone by aligning artwork themes and scale with the spatial and narrative affordances of the hosting interfaces and the designed themes of each zone. Detailed information about each zone, including themes and matching rationale, is shown in Table~\ref{tab:curatorial zones}.

\begin{table*}[t]
\caption{Curatorial zones of the MR Art Exhibition, organized by spatial logic, narrative and spatial rationale, representative artworks, and design elements.}
\label{tab:curatorial zones}
\centering
\resizebox{\textwidth}{!}{%
\begin{tabular}{@{}c p{2.2cm} p{2.4cm} p{3.5cm} p{3.5cm} p{3.6cm} p{3cm} p{2.5cm}@{}}
\toprule
\textbf{Zone\#} & \textbf{Location} & \textbf{Zone Name} & \textbf{Thematic Narrative} &
\textbf{Matching Rationale} & \textbf{Representative Artworks} &
\textbf{Supportive Design Elements*} & \textbf{Interaction Type} \\
\midrule

\textbf{1} & Student Activity Center (Indoor) & \textit{The Ocean of Origins} & “Trace the origins of life from the ocean, a journey to its very beginnings.” & 
Spatial: enclosed indoor; high-ceiling volume \newline Narrative: origin of life; tangible (object-based) & \textit{“AquaVerse”} (Figure~\ref{fig:Layout} A) \newline \textit{“The Rise of Species”} (Figure ~\ref{fig:Layout} C)& Exhibition logo (Figure~\ref{fig:Layout} B) & Clickable labels (Figure~\ref{fig:supportivedesign} B); voice interaction-"butterfly" (Figure~\ref{fig:supportivedesign} D)\\

\addlinespace[0.3cm]

\textbf{2} & Central Lawn – Right (Outdoor) & \textit{The Digital Garden} & “Digital lifeforms—plants, animals, and beyond—unfolding new visions of what life has become.” & 
Spatial: open outdoor volume; ground plane; building facades \newline Narrative: lifeforms; nature; tangible (object-based) & \textit{"Mycelial Memories"} (Figure~\ref{fig:Layout} D)\newline \textit{"Post-Nature of Growth"} (Figure~\ref{fig:Layout} E)& Directional arrows; directional sound & Clickable labels; voice interaction-"butterfly" \\

\addlinespace[0.3cm]

\textbf{3} & Front Plaza (Outdoor) & \textit{The Symbiotic Realms} & “A hybrid realm, serving as a dynamic stage for the exploration of wisdom, thought, and culture.” & 
Spatial: crossroads; open outdoor volume; ground plane \newline Narrative: culture; wisdom; intangible (concept-driven) & "Shared Sparks" (Figure~\ref{fig:Layout} G)\newline \textit{"Phantom Ark"} (Figure~\ref{fig:Layout} H) & Portal~1 (to a 360° VR campus environment) (Figure~\ref{fig:supportivedesign} C)& Clickable labels; voice interaction-"butterfly" \\

\addlinespace[0.3cm]

\textbf{4} & Central Lawn – Left (Outdoor) & \textit{The City of Illusions} & “Myth-reborn futures reimagined in a world shaped by uncertainty.” & 
Spatial alignment: grounds; building facades \newline Narrative alignment: Futures; imaginations & \textit{"Degenerative Utopia"} (Figure~\ref{fig:Layout} I)\newline \textit{"Nova Earth Odyssey (feat. Magu)"} (Figure~\ref{fig:Layout} J) & Directional sound; Portal~2 ( to a 360° VR future city environment) & Clickable labels; voice interaction-"butterfly" \\

\bottomrule
\end{tabular}%
}


\vspace{0.5em}
\raggedright
\footnotesize
*Supportive design elements applied across all zones include navigational signs (Figure~\ref{fig:supportivedesign} A), the "butterfly companion" and clickable labels showing artworks' introduction. \\
\end{table*}

\begin{figure*}[h]
  \centering
  \includegraphics[width=0.9\textwidth]{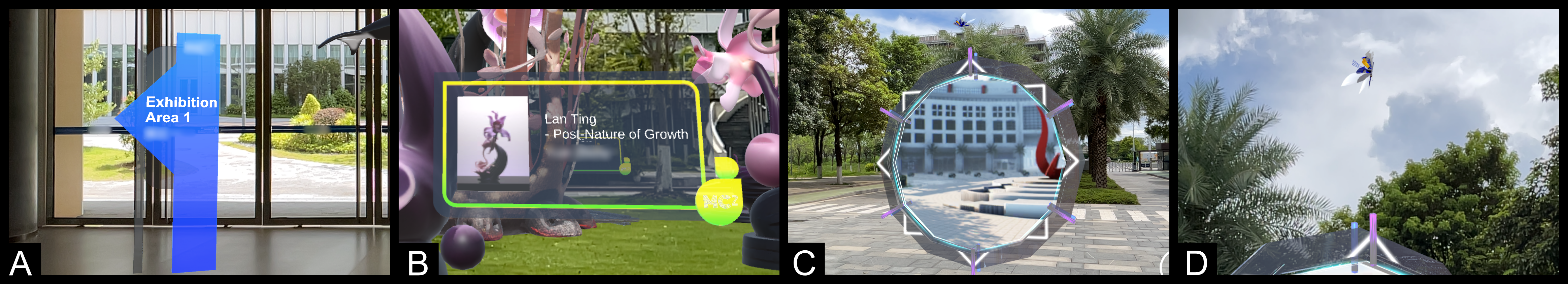}
  \caption{Supportive design elements (A-D). A shows one of the navigational signs indicating the exhibition area number and direction; B shows an information label with artwork details; C is one of the mirror-like "portals" to a VR environment; and D shows the "butterfly" companion.}
  \Description{Four thumbnails shows the supportive design elements (A-D).}
  \label{fig:supportivedesign}
\end{figure*}

Additionally, supportive design elements were introduced as supplementary exhibition features. Directional markers were added to provide visitors with navigational hints, clickable information labels offered layered details (e.g., names and descriptions of each artwork), and mirror-like “portals” were introduced to add a VR effect to the MR exhibition (A, B, \& C in Figure~\ref{fig:supportivedesign} respectively). A virtual LLM-driven "butterfly companion" (D in Figure~\ref{fig:supportivedesign}) functioned as both a conversational guide and an ambient presence. Directional sound cues were employed to support the localization of audio-based works.

\begin{figure*}[h]
  \centering
  \includegraphics[width=0.9\linewidth]{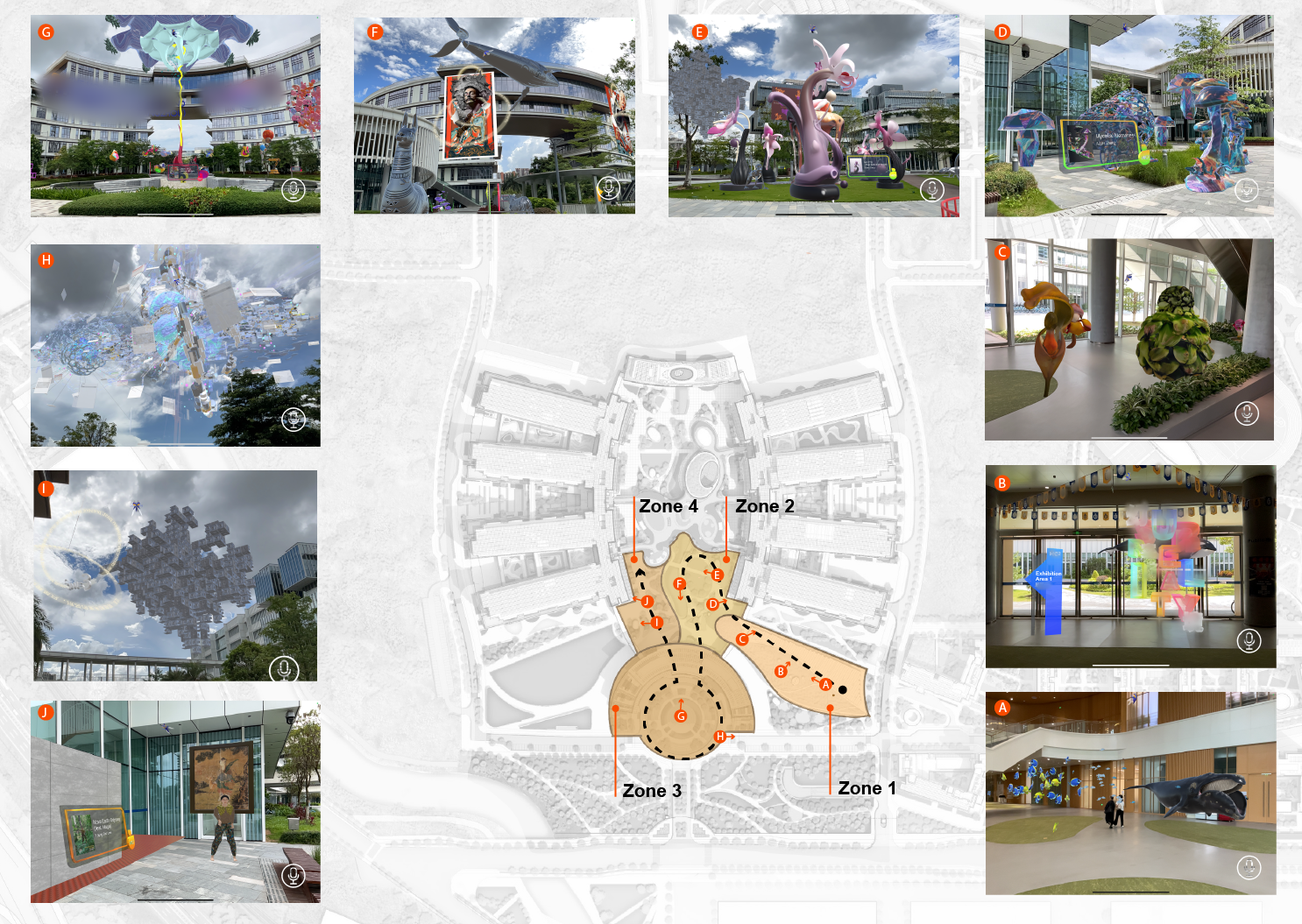}
  \caption{Layout of the MR art exhibition across four themed zones (Zone 1–4). The central map illustrates each zone’s footprint and anchor points of representative digital artworks, while the surrounding letter-coded thumbnails show representative headset views from corresponding locations, demonstrating how virtual works were integrated into the physical site.}
  \Description{An exhibition map diagram showing the footprints of four adjacent zones as viewing sequence. Zone 1 (indoor) is on the right, Zone 2 on the right-middle, Zone 3 at the bottom, and Zone 4 on the left-middle, together occupying the core area of the campus.}
  \label{fig:Layout}
\end{figure*}

As a result, the final exhibition covered approximately 26,000 square meters and was organized into four connected zones. Zone 1 started at the indoor Student Activity Center, Zone 2 was embedded in the right half of the open movement lawn area, Zone 3 was the front plaza, serving as a symbolic gateway, and Zone 4 was located on the left side of the central lawn. Figure \ref{fig:Layout} shows the visit routes with representative artworks, where visitors would encounter MR works in a gradual shift from everyday campus activity scenes to more symbolic and speculative installations at the institutional frontage, and then return to reflection.

\subsection{Technical Implementation}

Technical implementation translates the curatorial design into a functioning MR exhibition that can be reliably experienced on site and used as a research probe. This stage structures the technical work into an workflow that captures the physical environment, aligns virtual content with the reconstructed site, and deploys the exhibition across devices in the campus setting.

The implementation of the MR exhibition followed a three-stage workflow \cite{Numan_2025_OutdoorAugmentedRealityExperiences}: (1) \textit{capturing the site} by scanning and reconstructing the physical environment, (2) \textit{developing the exhibition} in Unity, and (3) \textit{deploying the application} to devices with on-site debugging \cite{10.1145/3746237.3746309}. Underpinning this workflow was a SLAM-based localization system \cite{alsadik2021simultaneous}, which ensured that virtual content was accurately anchored within the physical environment. The system operated on a server, receiving camera images transmitted from the device and returning positional data for rendering content in the Unity-based exhibition application.

In the first stage, the proposed spaces were scanned, and the collected data were processed to generate the information required for localization and to reconstruct a point cloud for subsequent Unity development, providing an accurate substrate for the MR exhibition. In the second stage, virtual artworks were optimized for performance, imported into Unity, and anchored to the reconstructed point cloud based on the curatorial design, ensuring precise alignment between digital overlays and their physical counterparts. In the final stage, the exhibition was packaged and deployed to prepared headsets, followed by an iterative cycle of on-site testing and refinement to ensure performance quality before release to public audiences.

This approach provides a transferable model for large-scale, site-specific MR exhibitions in outdoor public spaces, ensuring that the visitor experiences in subsequent studies are built on a stable and accurate MR system.

\section{Study 1: Focus Group with Experts}
\label{Section5_Study1}
\subsection{Methods}
\label{Section5_Study1_Method}
\subsubsection{Design and Objectives.}

To develop a professional and comprehensive understanding of opportunities, challenges, and design strategies involved in MR exhibition design and curation, we conducted a focus group with experts from different disciplines. Inspired by Richards et al.  \cite{Richards_2021_CoordinationChildrensBehavioralHealth}, who demonstrated how online focus groups can facilitate multi-stakeholder reflection and consensus-building, our session brought together experts from different disciplines to generate design feedback and foster critical reflection across perspectives. 

Our focus group had three goals: (1) to gather experts’ feedback on the MR exhibition from different professional perspectives; (2) to identify design strategies that worked well in our exhibition and to highlight strategies with potential for future development based on experts’ insights and past experience; (3) to discuss how specific MR design elements—such as narrative cues and sensory features—may support public engagement and immersion.

\subsubsection{Expert Participants.}

We purposively recruited eight experts across four domains—curation, architecture/urban design, art, and MR development/technology. These fields each address a different layer of MR exhibition making: curators guide narrative and audience experience, designers shape spatial conditions, artists define creative intent, and technologists support system stability and interaction. All invitees had previously visited our exhibition. Curators, artists, and technologists each reported 5+ years of MR exhibition experience. The two architects/urban designers had limited hands-on MR experience but were familiar with real-world MR exhibition cases and expressed a strong interest in the medium. All participants were based in Mainland China or Hong Kong.

\begin{table}[ht]
\centering
\footnotesize
\caption{Demographic information of expert participants in the online session.}
\label{tab: experts demographics}
\begin{tabular}{@{}l l c c c@{}}
\toprule
\textbf{ID} & \textbf{Role} & \textbf{Gender} & \textbf{Age} & \textbf{Years of Experience} \\
\midrule
C01 & Curator              & Male   & 36 & 10 \\
C02 & Curator              & Female & 35 & 11 \\
D01 & Landscape Architect  & Female & 29 & 6  \\
D02 & Architect            & Male   & 41 & 17 \\
A01 & Artist               & Male   & 32 & 9  \\
A02 & Artist               & Male   & 31 & 7  \\
T01 & Technologist         & Male   & 34 & 9  \\
T02 & Technologist         & Male   & 29 & 5  \\
\bottomrule
\end{tabular}
\end{table}

\subsubsection{Procedure.}
\begin{figure}[h]
  \centering
  \includegraphics[width=\linewidth]{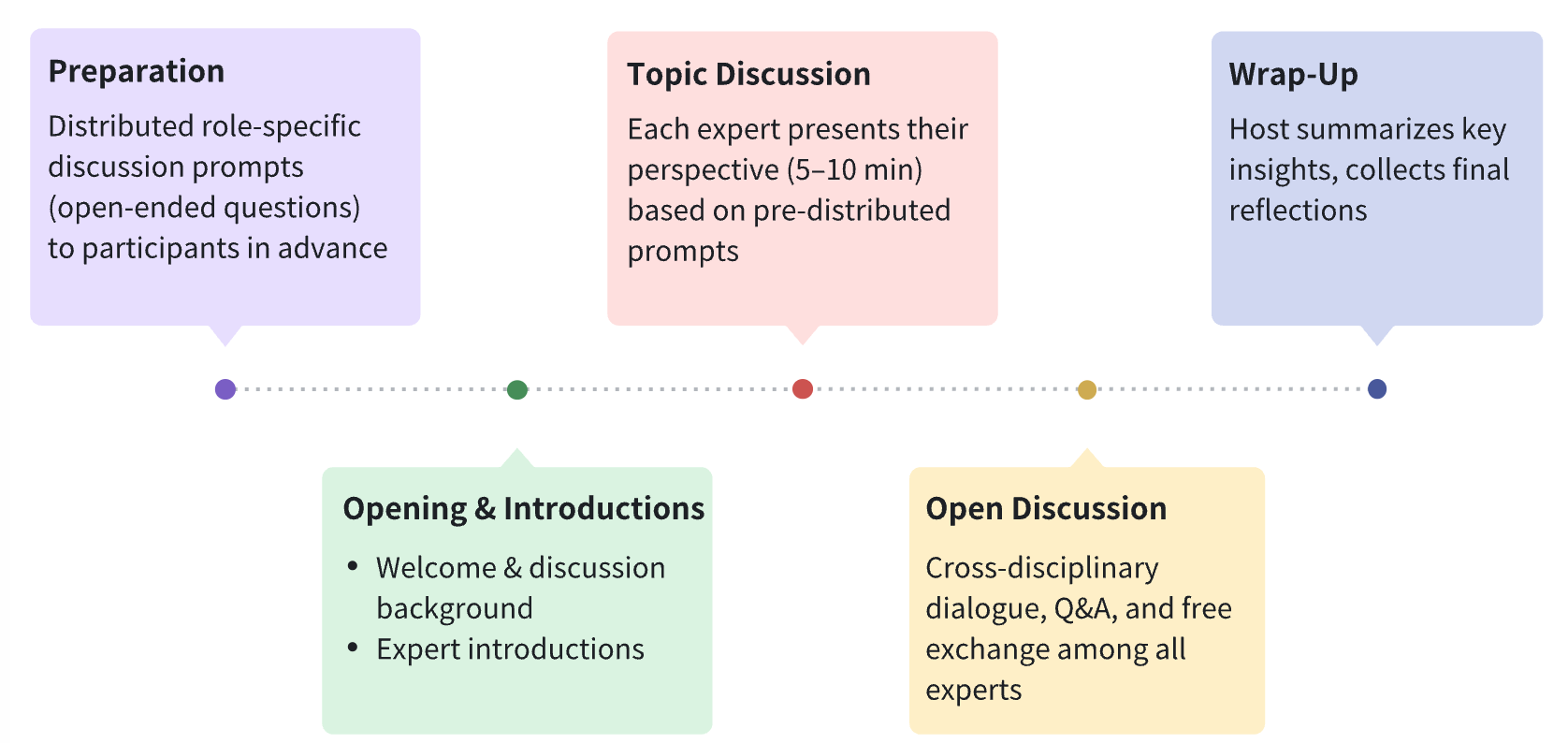}
  \caption{Structure of the 90-minute online focus group.}
  \Description{Structure of the online cross-disciplinary focus group, consisting of five sections: Preparation, Opening & Introduction, Topic Discussion, Open Discussion, and Wrap-Up.}
  \label{fig:90mStructure}
\end{figure}

The focus group consists of 5 steps (Figure~\ref{fig:90mStructure}). We prepared and distributed role-specific prompt questions (one week in advance) to guide discipline-grounded sharing of opinions. Each expert received a set of role-specific discussion prompts in the form of open-ended questions (Table \ref{tab:expert_prompts}). These prompts were tailored to their professional background and linked to the study’s research focus, including perspectives on the challenges and opportunities of MR exhibitions, reflections on their experience with the presented MR work, and their prior involvement with MR art or exhibition projects. The prompts were not meant to be answered beforehand; rather, they served as cues to help participants frame their viewpoints and elaborate on them during the session through individual reflections and group discussion.

The 90-minute online focus group was held on Tencent Meeting. It began with a short opening segment (10 minutes) where the host welcomed participants, introduced the project background, and invited each expert to briefly present themselves. The main body (about 60 minutes) consisted of panel discussions, in which the host sequentially cued one expert from each discipline to elaborate on the pre-distributed discussion prompts (5–10 minutes each), followed by an open discussion (15 minutes) that encouraged free exchange, Q\&A, and cross-disciplinary comments. The session concluded with a brief wrap-up (5 minutes), during which the host summarized key insights and invited final reflections on the focus group.

\subsubsection{Materials and Data Analysis.} \label{subsubsec:expert data analysis}
The focus group was hosted on Tencent Meeting, which supported synchronous video conferencing and screen sharing. Mentimeter was used to collect short text responses and instant feedback when experts reacted to one another’s statements, ensuring that all participants could contribute regardless of speaking order. During the session, the host shared slides that outlined the project background, the guiding topics, and the sequence of activities. All discussions were recorded with participants’ consent, and anonymized notes were taken by the research team to capture key points and illustrative quotes.

The collected materials included (1) the full video recordings from Tencent Meeting, (2) Mentimeter responses, and (3) the researchers’ observation notes. We conducted a qualitative thematic analysis following Braun and Clarke’s six-phase approach \cite{Braun_2006_psychology}. Two native Mandarin-speaking authors first familiarized themselves with the recordings and notes. One author then produced an initial set of 237 codes, capturing experts’ perspectives on design strategies, challenges, and opportunities in MR exhibitions in relation to our research questions. All the codes were translated into English and then reviewed and refined by the second author to ensure accuracy. Subsequently, 12 clusters were formed by identifying recurring semantic patterns across the codes, such as “space as story structure,” “physical–virtual fusion,” and “unique walking-based experience.” Through iterative discussion between the two authors, these groups were further clustered into four higher-level themes that reflected recurring cross-disciplinary concerns. The final themes were selected because they (a) demonstrated cross-disciplinary relevance, (b) appeared consistently across the dataset, and (c) aligned with our research questions.

\subsection{Results and Findings} \label{subsec:expert_results}

\subsubsection{Overall Impression.}

Across curatorial, design, artistic, and technical backgrounds, experts shared a generally positive and cohesive impression of our MR exhibition. Several referred to it as “a very good exhibition” and “a very impressive experience,” emphasising that it exceeded their initial expectations despite many of them arriving “without a clear mental model” of what such an MR show would look like. Most described the exhibition as highly immersive, noting that they were able to engage with the MR experience quickly despite the unfamiliar scale and format. The large indoor–outdoor spatial configuration was consistently viewed as distinctive, and several experts highlighted that the walking-based format enabled a smooth transition into the mixed-reality environment.

Experts also reported a strong sense of presence and spatial coherence. Moving between zones felt intuitive, and the digital overlays blended naturally with the campus environment. Overall, they described the exhibition as engaging, visually compelling, and conceptually promising, with many expressing enthusiasm about its experimental value and future potential.

Some critiques also emerged. Several experts noted accessibility concerns related to the technical setup, such as the discomfort of wearing a headset or the possibility of dizziness—issues largely tied to current hardware limitations. Others mentioned safety concerns: while the campus setting felt manageable, similar MR experiences in busier public spaces might create risks related to crowd movement. Despite these points, the overarching impression was highly positive. Experts agreed that the exhibition demonstrated the potential of large-scale MR deployments and offered a convincing glimpse into how mixed-reality experiences may develop in the future.

\subsubsection{Immersive Spatial Experience.}

Experts described the MR experience as strongly immersive, emphasising how scale, movement, and environmental cues shaped their engagement. Several noted that they became absorbed in the mixed-reality environment more quickly than expected. As one curator reflected, \textit{“there were many aspects that exceeded my expectations, yet they still felt coherent afterwards”} (C01). Technical experts likewise framed the experience as a forward-looking demonstration, with one describing it as \textit{“a glimpse of what the future may look like”} (T01).

Walking emerged as a defining experiential element. Rather than remaining in a single controlled room, participants moved continuously through indoor, outdoor, and transitional spaces. A curator described this as \textit{“a walking state integrated with the real environment”} (C01), while another emphasised the rarity of encountering MR installations that unfold across such a large physical area (C02). This mobility created a sense of progressive revelation; as C01 noted when walking through Zone 2, it felt like “entering a virtual garden, where each step offered a new scene.”

Experts also emphasised the role of multi-sensory cues in deepening immersion. One artist recounted a moment on a lakeside pathway (in Zone 4) when they suddenly heard \textit{“the sound of an insect passing by my ear”} and briefly questioned whether it was part of the MR system or the real surroundings (A02). The combination of ambient sound, natural airflow, and environmental scent with digital overlays was repeatedly cited as a key factor in dissolving the boundary between physical and virtual elements.

The spatial scale further amplified immersion. For an urban design expert experiencing MR headsets in the outdoor space for the first time, the exhibition felt \textit{“very shocking,”} particularly because the MR installations \textit{“achieved scales that traditional means could never reach”} (D01). Technical specialists noted that certain long-distance views produced a heightened sense of volumetric realism, with one observing that AI-generated scenes placed against buildings in Zone 2 (see Figure \ref{fig:facade artwork}) appeared \textit{“almost three-dimensional”} from afar (T02).

The experts believed that immersion did not stem solely from digital visuals. Instead, it was created through movement, the scale of the spaces, the atmosphere, and the integration of different senses. Their reflections indicate that placing MR artworks in real-world environments is crucial for creating a coherent and engaging experience.

\subsubsection{Narrative Structuring in MR Space.} \label{subsubsec:narrative}

Experts described the exhibition’s narrative as something that unfolded through the act of moving, looking, and inhabiting the mixed environment, rather than through any explicit storyline. They emphasised that meaning accumulated incrementally as visitors progressed across sites, with spatial transitions functioning as narrative pivots. One curator noted that the sequence of scenes created \textit{“a feeling of being guided into different worlds without being told explicitly what story I was entering”} (C02). This sense of tacit narrative framing was repeatedly mentioned as a distinctive quality of the experience. 

Several experts observed that narrative emerged through the emotional and atmospheric shifts embedded in each zone. For example, entering areas with underwater creatures or multispecies imagery was described as \textit{“immediately signaling a particular emotional orientation”} (C02) that established the exhibition’s thematic trajectory. Experts interpreted this not merely as an aesthetic choice but as an intentional narrative gesture that encouraged reflection on ecological relations. One expert explained that such motifs invited audiences to consider \textit{“how humans coexist with non-human others, and what kinds of awareness this might activate”} (C01). This showed that experts were interpreting spatial changes as narrative cues.

While two experts (C02 and A01) emphasized that movement should play a decisive role in how the narrative is perceived, they also pointed out that the current exhibition did not yet fully consider the bodily transitions that shape narrative understanding. Experts repeatedly described walking as a way of revealing story elements—turning a corner, crossing a bridge, or approaching a structure often triggered emotional or conceptual shifts that could be more intentionally integrated. As one expert noted, \textit{“the story should changes when your body changes direction; the narrative is in the way you move”} (T02). This perspective highlights a broader view of narrative in MR: it is not simply delivered to the viewer but discovered through the subtle, moment-to-moment movements of an embodied trajectory.

\subsubsection{Technological Conditions of MR Curation.}
Experts noted that the technical system shaped both the strengths and limits of the MR experience. Many described the spatial scanning and anchoring pipeline as the foundation that made the exhibition possible. One technical expert explained that \textit{“everything depends on a deep scan of the space; without this database, nothing else would hold together”} (T01). This stability gave the exhibition its large-scale coherence.

At the same time, experts pointed to clear constraints. Outdoor environments raised concerns about tracking reliability, safety, and changing light. As one expert put it, \textit{“outdoor MR always negotiates safety and weather before it can negotiate creativity”} (T02). These conditions shaped which movements, directions, or viewing angles were feasible in practice. Rendering performance was another issue. Some experts observed a visible gap between artistic intention and device capability. One artist described moments when the digital elements produced \textit{“a grain that breaks the illusion”} (A01). Another noted that limited computational power sometimes simplified textures or lighting, resulting in \textit{“a reminder that the virtual is still catching up with the real”} (A02). These comments show how technical limits influence atmosphere and realism.

Technology also affected authorship. When certain effects could not be rendered, technical teams had to reinterpret or adjust works. One expert explained that \textit{“sometimes the artwork becomes what the system can support, not what the artist imagined”} (T01). While this could be restrictive, some experts also saw creative potential in such limitations. A curator suggested that \textit{“technical limits might become narrative tools of their own”} (C01).

In addition, experts expressed concerns about how technology shapes accessibility. They noted that immersive MR experiences can bring audiences “\textit{closer to the artwork}” because the pieces appear to unfold directly around them (A01). However, they also emphasized that dependence on head-mounted displays imposes operational constraints. As one curator revealed, “\textit{running an MR exhibition like this can be difficult, and the available devices limit how many people can experience it}” (T02). This tension highlights the dual nature of MR technology: it enhances immediacy and engagement, yet also restricts scalability and public reach due to hardware cost and availability.

\section{Study 2: User Study }\label{sec:user study }
\label{Section6_Study2}
\subsection{Methods}
\label{Section6_Study2_Method}
\subsubsection{Study Design and Objective. }
In this user study, our objective was to understand how general audiences evaluated the MR art exhibition and how the exhibition’s design shaped visitors’ experiences. Following established guidance on mixed-methods research in HCI \cite{lazar2017research}, we employed both questionnaire and semi-structured interviews, combining the breadth of questionnaire data with the depth of qualitative inquiry. The questionnaire was designed to capture characteristics of a larger sample of general visitors than would have been feasible through interviews alone, as well as their baseline attitudes and initial reactions to the exhibition. The interviews aimed to gain deeper insight into participants’ specific perceptions of the exhibition’s design and their overall experience, and to elicit more nuanced reflections on how the MR elements influenced their engagement.

The questionnaire comprised three parts. The first part consisted of single-choice questions about participants’ \textit{background information}, including demographic characteristics, prior experience with extended reality, art, or design, and familiarity with the campus environment. The second part used Likert-scale items to assess \textit{overall impressions} \cite{rokhsaritalemi2020review,Bareisyte_2024_realitysystematicscopingreview} of the exhibition, including the perceived integration between exhibition content and physical space, how the MR format shaped their experience of the artworks and space, and the extent to which the exhibition maintained their attention throughout the experience. The third part used Likert-scale ratings of \textit{specific experiential dimensions}—perceived engagement with two items \cite{doherty2019EngagementHCI}, immersion with three items \cite{bowman2007virtual,shin2019HowDoes}, sense of presence with two items \cite{tran2024SurveyMeasuring}, and evaluation of visual aesthetics with two items \cite{bhandari2019UnderstandingImpact}—as well as participants’ intention to engage with similar MR exhibitions in the future \cite{Kim_2023_ExpertSupportSystemArt}.

The semi-structured interviews were guided by eight open-ended questions designed to elicit in-depth reflections on the experiential quality of the exhibition (see Appendix~\ref{User_Questionaires}). These questions extended and deepened the themes from the questionnaire to ensure coverage of key topics. Core prompts invited participants to describe their overall impressions of the exhibition, the aspects they found most memorable and why, how the MR format influenced their experience of the space and artworks, how they perceived the integration of digital content with the physical environment, and their feedback and suggestions on the exhibition design. Throughout the interviews, interviewers also asked follow-up questions to probe for further details and concrete examples.

\subsubsection{Participant Recruitment. }

Participants were recruited from members of the general audience who registered for and attended the MR art exhibition. Registration for the exhibition was advertised through the university’s official WeChat public account and was open to the general public. In total, 71 participants completed the questionnaire. Most were between 26–35 years old (43.7\%) or 36–45 years old (38.0\%), with smaller groups aged 18–25 (4.2\%), 46–55 (8.5\%), and over 55 (5.6\%). Additional demographic information—including campus familiarity, prior exposure to extended reality, backgrounds in design or art, and digital art knowledge—is reported in Table~\ref{tab:audience demographics} in Appendix~\ref{Apex:Demographic profile of the participants (N=71)}. From this questionnaire sample, a subset of participants was randomly invited to take part in the follow-up semi-structured interview; in total, 21 participants agreed and were interviewed. Details of the interview participants are provided in Table~\ref{tab: Interview Participants} in Appendix~\ref{Apex:List of Interview Participants}.

\subsubsection{Procedure.}
Visitors registered for specific 30-minute exhibition slots, organized as guided tours. Each slot accommodated up to five registered general visitors and was led by a trained tour guide who ensured safety, provided explanations, and responded to questions. After viewing the exhibition, visitors who consented to take part in the study (hereafter, participants) were invited to complete the questionnaire. Those who had additionally volunteered for the interview were then randomly selected and interviewed in sessions lasting 30–40 minutes. The interviews followed a semi-structured protocol based on eight open-ended questions designed to elicit in-depth reflections on the experiential quality of the exhibition (see Appendix~\ref{User_Questionaires}). Most interviews were conducted individually, while three involved pairs of participants.

\subsubsection{Materials and Data Analysis. }
Questionnaire data were collected through a self-administered online questionnaire developed using Microsoft Forms, which is widely accessible and familiar to participants in the study context. This platform supports customizable questionnaires and the export of raw data in structured formats.

The data collected in this study consisted of (1) questionnaire responses derived from Microsoft Forms and (2) audio recordings and transcripts of the interviews. Questionnaire data were analysed quantitatively using descriptive statistics (frequencies, percentages, means, and standard deviations) to provide an overview of participants’ responses to each item. Interview transcripts were analysed qualitatively using thematic analysis following Braun and Clarke’s six-phase approach \cite{Braun_2006_psychology}, employing the same analytic procedure as in study \ref{subsubsec:expert data analysis}. We generated 204 initial codes capturing participants’ evaluations of how the design shaped their experience of the exhibition. After being reviewed by a second author, the initial codes were grouped into 20 code groups based on semantic similarity, co-occurrence, and relevance to the research questions. Subsequently, four higher-level themes were developed through iterative discussions among the authors, and these are reported in Section \ref{subsubsec:Positive Overall Impressions} through Section \ref{subsubsec:design interventions}.

\subsection{Results and Findings}

\subsubsection{Questionnaire Results. }
\begin{figure*}[h]
  \centering
  \includegraphics[width=0.8\linewidth]{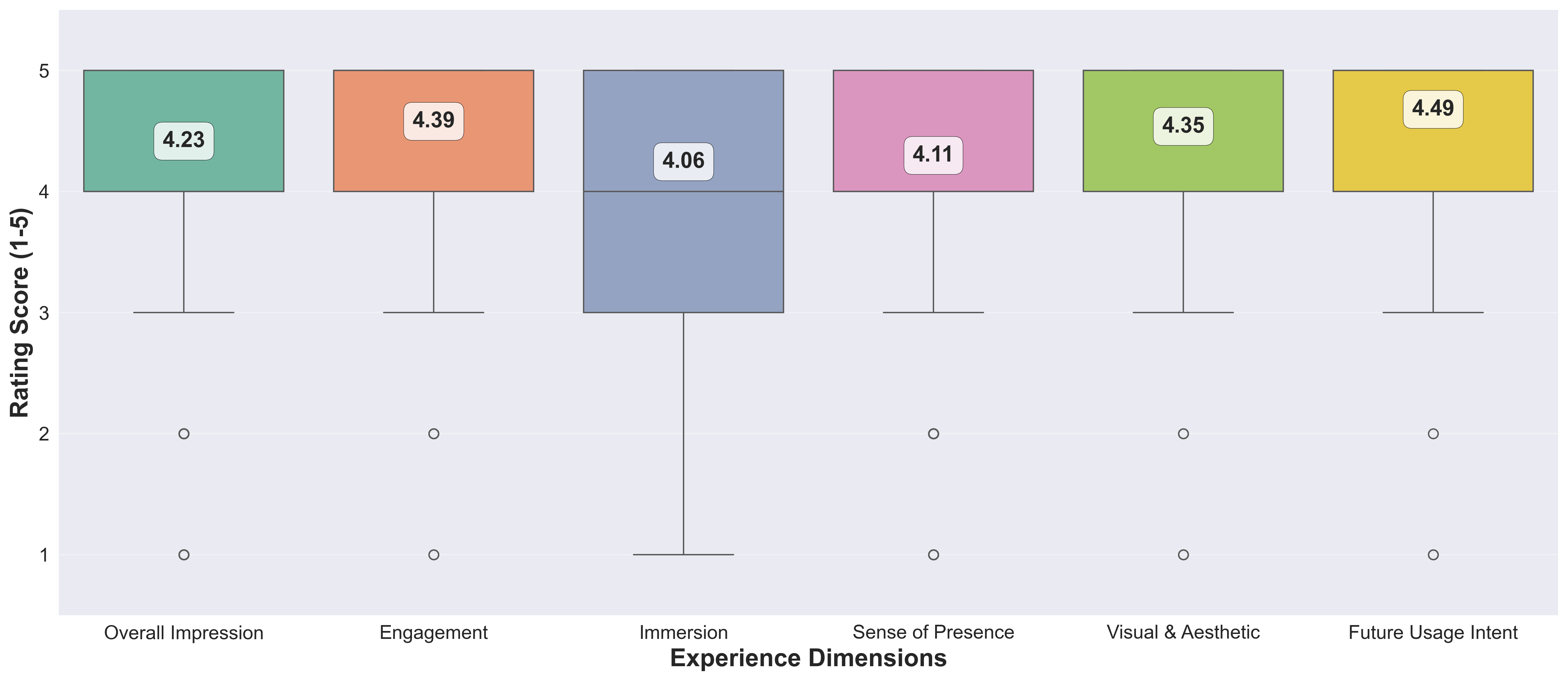}
  \caption{General user experience ratings across six key aspects of the MR art exhibition (Mean values shown above each box).}
  \Description{A chart shows boxplots of audience experience ratings across six dimensions of the MR exhibition. The boxplots communicate the mean, interquartile range, standard deviations, and outlier values for each dimension.}
  \label{fig:boxplot}
\end{figure*}

The questionnaire results indicate that participants generally reported positive experiences across all six dimensions of the MR exhibition, with mean scores consistently above 4.0 on a 5-point scale (Figure \ref{fig:boxplot}). All six dimensions demonstrated acceptable to excellent internal consistency, with a Cronbach’s $\alpha$ ranging from 0.78 to 0.91 (see Table \ref{tab:survey items}).

The \textit{Overall Impression} was rated highly (M = 4.23, SD = 0.80), suggesting that the integration of virtual and real elements was perceived as smooth and engaging. \textit{Engagement} (M = 4.39, SD = 0.88) and \textit{Visual \& Aesthetic Evaluation} (M = 4.35, SD = 0.89) were among the strongest dimensions, indicating that participants felt actively involved in the interactive elements and found the aesthetic qualities of the MR exhibition compelling.

By contrast, \textit{Immersion} (M = 4.06, SD = 0.86) and \textit{Sense of Presence} (M = 4.11, SD = 1.03) received slightly lower ratings, though still positive overall. These results suggest that while the exhibition successfully created a sense of presence and immersion, there is room for improvement in strengthening the seamlessness of the virtual–real blend and the depth of immersion.
Finally, \textit{Future Usage Intent} achieved the highest rating (M = 4.49, SD = 0.79), demonstrating that participants were highly willing to recommend the MR exhibition to others and expressed interest in experiencing similar applications in the future. This finding highlights the perceived value and scalability of MR exhibitions in public space contexts. The next sections will present the themes generated from interview.

\begin{table*}[h!]
\centering
\footnotesize
\renewcommand{\arraystretch}{1.2}
\caption{Questionnaire items organized by six key dimensions of the MR art exhibition experience, with their factor loadings and \emph{item-level} descriptive statistics (Mean, Std. Dev., Cronbach’s $\alpha$).}
\label{tab:survey items}
\begin{tabular}{@{}l p{6cm} c c c c@{}}
\toprule
\textbf{Considered Aspects} & \textbf{Items} & \textbf{Factor Loadings} & \textbf{Mean} & \textbf{Std. Dev.} &\textbf{Cronbach’s $\alpha$} \\
\midrule
\multirow{10}{*}{Overall Impression \cite{rokhsaritalemi2020review,Bareisyte_2024_realitysystematicscopingreview}}
  & • The virtual and real integration in this exhibition was smooth and seamless.
    & 0.85 & 4.07 & 0.80 & \multirow{10}{*}{0.85}\\
  & • It was easy to switch between virtual and real elements during the exhibition.
    & 0.82 & 3.93 & 0.95 \\
  & • The MR technology provided me with a new perspective on art appreciation.
    & 0.78 & 4.51 & 0.83 \\
  & • The MR technology enhanced my overall campus exhibition experience.
    & 0.79 & 4.45 & 0.82 \\
  & • The exhibition maintained my attention throughout the experience.
    & 0.76 & 4.20 & 0.90 \\
\addlinespace[0.3cm]

\multirow{4}{*}{Engagement \cite{doherty2019EngagementHCI}}
  & • I felt actively engaged in the interactive elements of the exhibition.
    & 0.88 & 4.28 & 0.88 & \multirow{4}{*}{0.83}\\
  & • The exhibition successfully sustained my attention and interest.
    & 0.84 & 4.51 & 0.79 \\
\addlinespace[0.3cm]

\multirow{3}{*}{Immersion \cite{bowman2007virtual,shin2019HowDoes}}
  & • I felt completely immersed in the virtual environment.
    & 0.86 & 4.20 & 0.86 & \multirow{3}{*}{0.86}\\
  & • I lost track of time while experiencing the exhibition.
    & 0.79 & 4.03 & 0.96 \\
  & • The virtual and real environments were well integrated.
    & 0.83 & 3.96 & 1.07 \\
\addlinespace[0.3cm]

\multirow{3}{*}{Sense of Presence \cite{tran2024SurveyMeasuring}}
  & • I felt a strong sense of presence in the blended virtual-real environment.
    & 0.89 & 4.13 & 1.03 & \multirow{3}{*}{0.78}\\
  & • I felt physically present in the exhibition space.
    & 0.85 & 4.10 & 1.06 \\
\addlinespace[0.3cm]

\multirow{4}{*}{Visual \& Aesthetic Evaluation \cite{bhandari2019UnderstandingImpact}}
  & • The visual elements of the exhibition were appealing and engaging.
    & 0.91 & 4.25 & 0.89 & \multirow{4}{*}{0.84}\\
  & • The MR technology enhanced the aesthetic impact of the art pieces.
    & 0.87 & 4.45 & 0.81 \\
\addlinespace[0.3cm]

\multirow{3}{*}{Future Usage Intent \cite{Kim_2023_ExpertSupportSystemArt}}
  & • I would recommend this MR exhibition to others.
    & 0.92 & 4.48 & 0.79 & \multirow{3}{*}{0.91}\\
  & • I would like to experience similar MR applications in the future.
    & 0.88 & 4.51 & 0.83 \\
\bottomrule
\end{tabular}
\end{table*}

\subsubsection{Positive Overall Impressions.} \label{subsubsec:Positive Overall Impressions}
In the interviews, 19 of the 21 participants reported positive overall impressions of the exhibition, while two expressed neutral attitudes. The strong positive response was primarily attributed to the novelty of experiencing a large-scale outdoor art exhibition in mixed reality for the first time, which created a sense of immersion distinct from traditional exhibition formats. As one participant noted, \textit{“The overall experience was very good, completely different from the audiovisual impressions I usually get from traditional exhibitions” }(P35). Another described it as \textit{“a visual experience like entering a game in real life”} (P71), while others highlighted the \textit{“powerful visual impact”} and the memorable combination of virtual and physical elements (P31, P32).

Participants emphasized that what made the exhibition stand out compared to conventional art shows was the integration of digital and physical environments, the interactive qualities enabled by this hybrid form, and the sense of immersion it created. For example, one participant explained: \textit{“The biggest difference is in the form of presentation—it directly confronts me with the content, unlike traditional art exhibitions where you just walk up to a static object and read its description” }(P19). Another reflected that, \textit{“Compared to traditional art exhibitions, this one feels more like I am part of the exhibition itself”} (P52). Similarly, one participant remarked, \textit{“Digital content appearing naturally alive on site, makes the experience much more immersive” }(P57).

\subsubsection{Technological Limitations Become Barriers to the Experience.}
Despite overall positive feedback, negative comments primarily highlighted discomfort during the MR exhibition tour due to technological limitations. First, direct body discomfort was caused by wearing headsets of a certain weight and their Wearing experience. As two participants noted, \textit{"The headset was a bit heavy for me to keep on for tens of minutes"} (P21) and \textit{"For my friend, it might feel like she can't stand it for five minutes, while others might want to keep playing even after this tour"} (P35). Additionally, there were indirect issues with the comfort of the equipment, as P35 mentioned: \textit{"I wear glasses. It was a bit inconvenient. At first, my glasses fogged up, so I had to take them off and wipe."} These discomforts may cause distractions from the whole exhibition experience, \textit{"The heaviness of the headset had taken some of my attention" }(P21).

Additionally, since the visuals participants see are digitally displayed in the headset, some participants considered issues with resolution or display quality, in comparison to reality, a barrier to the experience. As a participant put it \textit{"The headset doesn't offer the best depth perception, so the real environment sometimes doesn't feel as realistic compared to actual reality."}(P02) And \textit{"In the headset display, people in the real world sometimes appear intermittently with blurred edges, especially those in the distance, which caused some confusion."} (P65)

\subsubsection{Spatial and Coherent Integration as a Driver of Experience.}

Participants consistently emphasized that seamless integration was critical to producing a high-quality experience. They described three perspectives on integration, which can be grouped into two broad categories: the integration of digital artworks into the physical environment and the integration among digital artworks themselves.

\textbf{(1) Integration into the Physical Environment.}

\begin{figure}[h]
  \centering
  \includegraphics[width=0.9\linewidth]{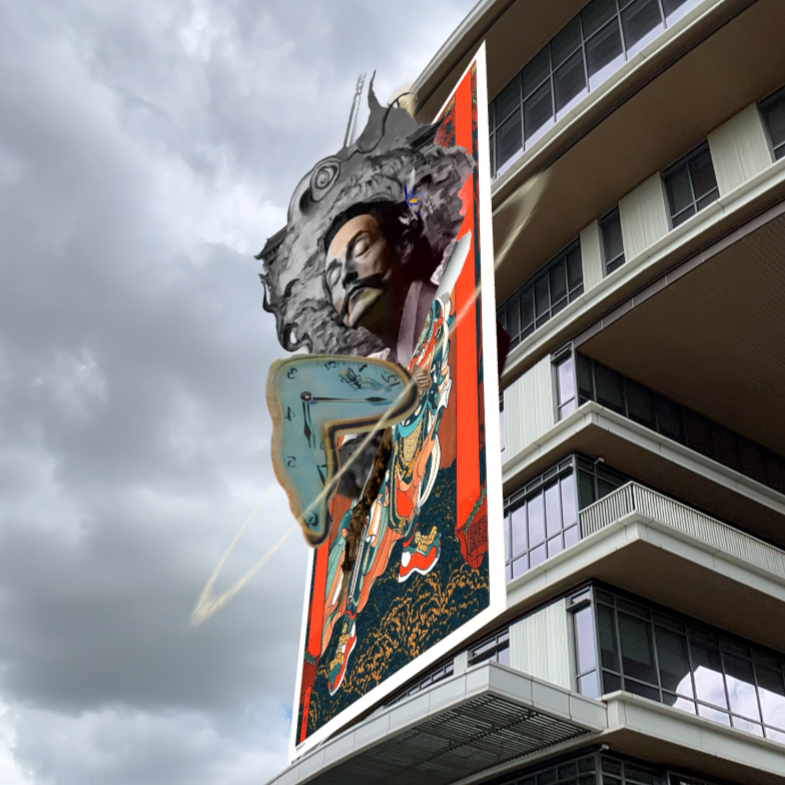}
  \caption{The representative artwork in zone 3, featuring a 3D digital poster of Dali’s Portrait displayed on the facade of a modern academic building.}
  \Description{A 3D digital poster of 'Dali's portrait' on the facade of the modern style academic building.}
  \label{fig:facade artwork}
\end{figure}

Participants highlighted two complementary forms of integration between digital artworks and physical space. The first was \textbf{\textit{spatial–physical alignment}}, where virtual artworks were precisely fitted to architectural surfaces or spatial features. The most frequently mentioned example was the giant 3D dynamic, poster-like artwork, which appeared as if it were pasted directly onto the building façade (Figure \ref{fig:facade artwork}). Participants consistently described this alignment as highly realistic and immersive. For example, one participant stated, \textit{“My top choice was the giant 3D poster on the building façade—it felt like it was truly pasted onto the wall”} (P52). Two others echoed this point, noting that \textit{“the coordination between the huge poster and the wall surface made the virtual content blend very tightly with the real space”} (P68, P69). Such comments illustrate how the \textit{“degree of spatial fit”} was perceived as a key factor enhancing both realism and immersion \cite{Jiang_2025_user-centreddigitalartdesign}. The positive responses suggest that this design consideration successfully contributed to participants’ sense of coherence and immersion.

\begin{figure}[h]
  \centering
  \includegraphics[width=0.9\linewidth]{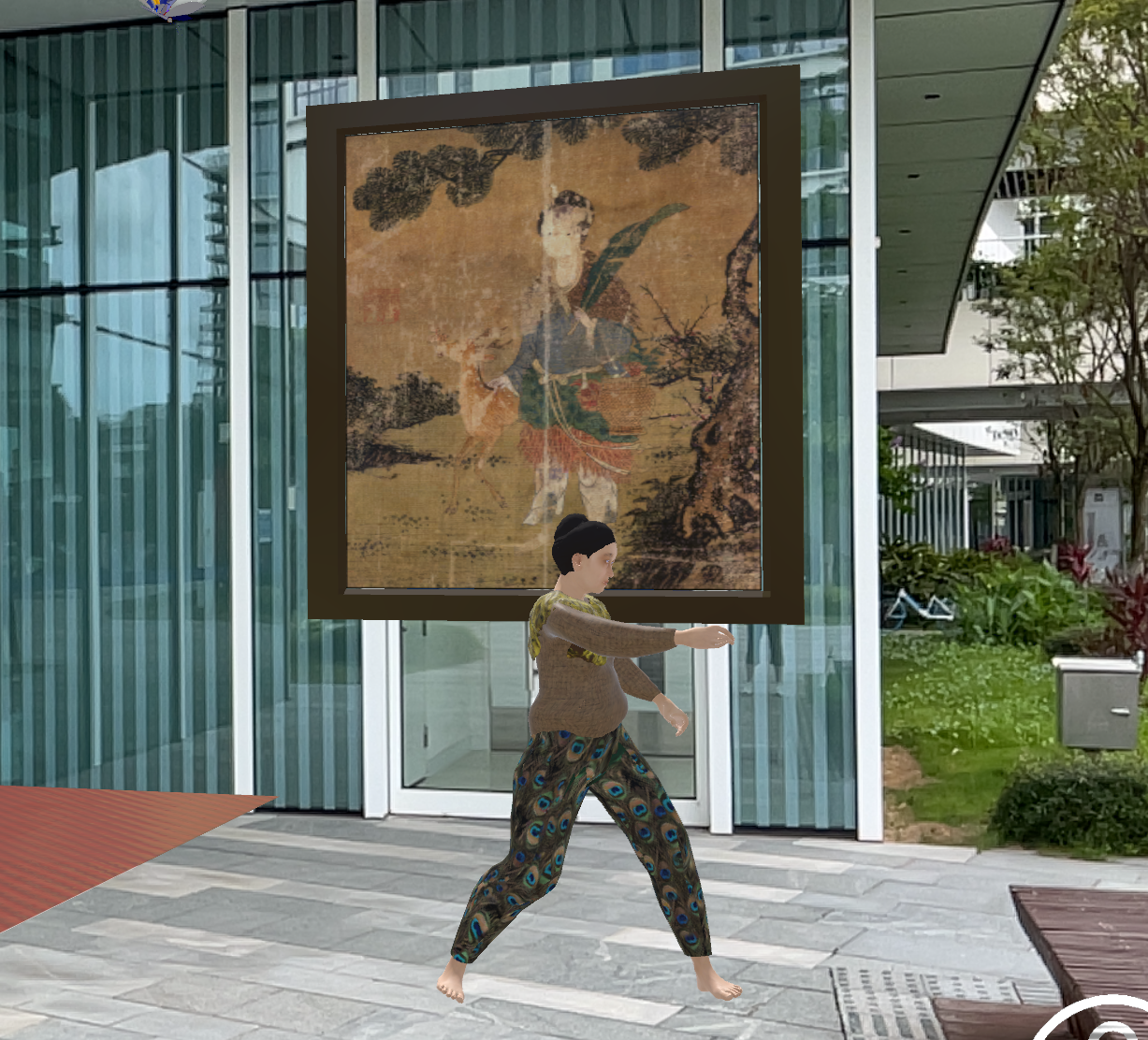}
  \caption{The representative artwork "Magu" in zone 4. With a 3D-modeled woman dancing in front of a Korean ancient drawing}
  \Description{A 3D-modeled woman dancing in front of a Korean ancient drawing.}
  \label{fig:dance}
\end{figure}

\begin{figure}[h]
  \centering
  \includegraphics[width=0.9\linewidth]{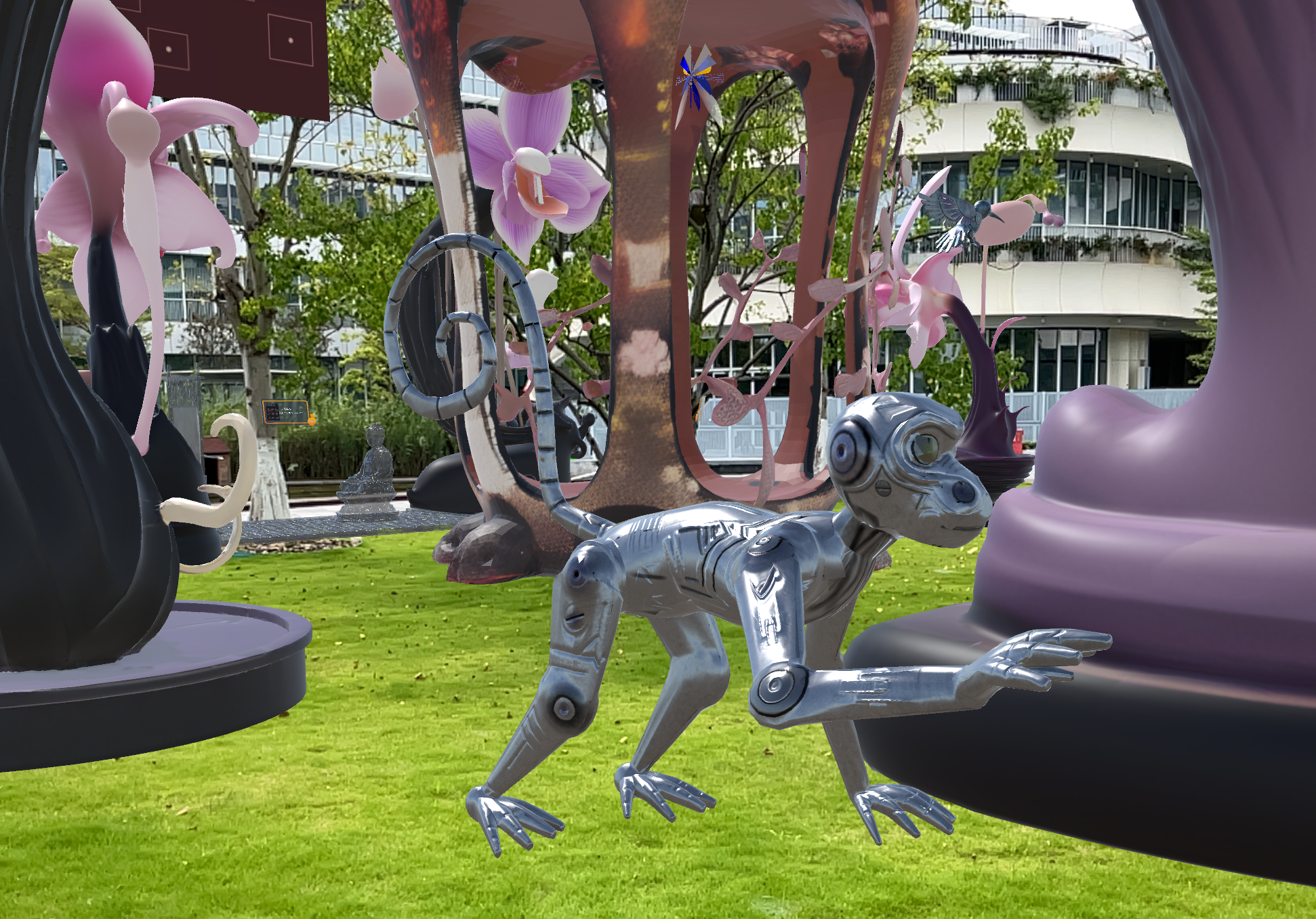}
  \caption{Animal- and plant-shaped artworks situated on the lawn.}
  \Description{A 3D-modeled monkey with mechanical textures located on lawn area. Also located on lawn area, are some 3D-modeled plants such as orchids on the background.}
  \label{fig:anmal-plant-shaped}
\end{figure}

The second form was \textit{\textbf{perceptual–logical coherence}}, which referred to whether digital elements felt natural and consistent when experienced in context. Participants explained that beyond technical alignment, what mattered was whether the content felt appropriate for the environment. Several pointed to a 3D figure of a dancer (“Magu”) as particularly effective (Figure~\ref{fig:dance}). They described how placing a human-like figure in the campus context felt ordinary, familiar, and therefore convincing. As one participant put it: \textit{“Magu was simple—it was just a person. Since I am also a person in this setting, it felt like she was living here with me”} (P57). Others agreed, adding that this sense of everyday relatability reduced the distance between the audience and the artwork (P68, P69). Similar impressions were noted in Zone 2, where animal- and plant-shaped artworks were situated on the lawn (Figure~\ref{fig:anmal-plant-shaped}). One participant explained: \textit{“The colorful mushrooms on the lawn and bushes felt natural—there was no sense of separation from reality”} (P66). Another highlighted how movement added to this harmony: \textit{“The metallic animal artworks on the grass felt harmonious.”} (P70). These examples illustrate that coherence was understood not only in terms of visual fit but also in terms of logical appropriateness to the site and everyday expectations.

\textbf{(2) Integration among Digital Artworks.}

\begin{figure}[h]
  \centering
  \includegraphics[width=0.9\linewidth]{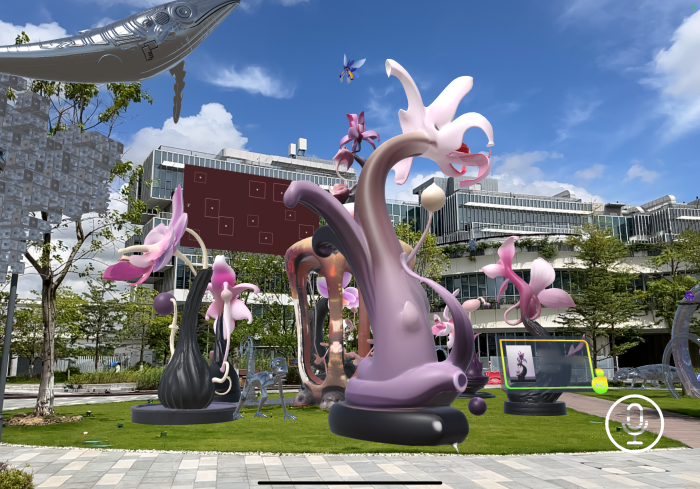}
  \caption{A representative artwork installed in Zone 2 (\textit{The Digital Garden}).}
  \Description{A Mixed-reality garden environment with digital content coherency.}
  \label{fig: garden}
\end{figure}

Beyond the relationship with physical space, participants also emphasized coherence among digital artworks themselves within the shared environment. This perspective concerned whether the layout, density, scale, form, and color of different works were arranged in ways that collectively created a harmonious atmosphere. One participant (P35) drew a vivid analogy to the aesthetics of a traditional Chinese garden, where human, natural, and environmental elements are interwoven \cite{zheng2024tourist,chen2009sustainable,keswick2003chinese}: \textit{“What I found most interesting was how the artworks did not feel independent from one another as in a traditional exhibition. Instead, they were layered together, creating a harmonious atmosphere like the rhythm of a Chinese garden, where every step reveals a new scene. Just as the pavilions, landscapes, and seasonal changes in a garden are always in dialogue, the artworks here are combined to form a coherent environment that made me feel part of it”}.
P02 also mentioned the balance between artworks: \textit{"The flower (garden) exhibition (zone 2) I found was super well balanced. The size was nice. The space between the pieces was really nice, and the arrangement was great."} (see Figure \ref{fig: garden}).

\subsubsection{Supportive Design Interventions that Captured Attention.}\label{subsubsec:design interventions}

Beyond the virtual artworks themselves, participants also highlighted several design interventions that significantly enhanced their sense of novelty, impression, and positive engagement with the exhibition.

 One of the most frequently mentioned elements was the MR-to-VR transition “portal,” designed as a mirror-like gateway placed at two points in the exhibition (thumbnail C in Figure~\ref{fig:supportivedesign}). Stepping through the portal symbolized moving from a mixed reality space into a fully virtual one, producing a strong sense of spatial teleportation. Participants described the experience of shifting from a visually grounded environment into a fully digital setting as psychologically impactful. For example, one participant reflected, \textit{“I found the portal very interesting. Moving back and forth between the real and the virtual gave me imaginative ideas...” }(P31). Another recalled the strong sense of realism during the transition: \textit{“The portal left a deep impression. The moment of switching into VR felt very real...It made me nervous—I hesitated to take the first step from one space to the other.”} (P65). Similarly, one participant described the portal as a “black hole”: \textit{“Once inside, it became a 360-degree VR environment. I suddenly felt like I had slipped from the real world into a purely virtual one...”} (P71).

Another design element that attracted strong attention was the use of sound effects. Multiple participants noted that the sound enhanced the overall atmosphere, transforming the exhibition into a multisensory experience. As one participant explained, \textit{“Sound was something I really noticed...being immersed in a soundscape made me feel more part of the exhibition atmosphere”} (P66). Spatialized sound also directed participants’ attention toward specific works: \textit{“I think directional sound was used here. As I walked into a space, I suddenly heard something from one direction, which made me turn and look...” }(P52).

Finally, participants highlighted the role of the virtual guide element, represented by a butterfly that accompanied them throughout the experience. The guide functioned as a subtle interface, simultaneously providing a sense of psychological comfort and reminding participants of system boundaries to prevent over-immersion. As one participant noted, \textit{“I kept noticing a butterfly following me, and whenever I looked back and saw it there, I felt secure”} (P49).

\section{Overall Discussion }

In this section, we synthesize insights from the design process, focus group, and user study to discuss how our findings extend prior understandings of the constraints and possibilities of MR exhibitions in public space. We also outline strategies that can guide future public-space MR exhibition practices and explain how contextualism operates as a curatorial and design principle for embedding digital layers within existing environments. Finally, we reflect on the limitations of our design and study to highlight directions for future research and practice.

\subsection{MR Exhibition Curation Opportunities and Challenges (RQ1) }

Our MR Art Exhibition revealed several opportunities and challenges in curating immersive experiences within existing public environments. Drawing from our full pipeline—from contextual site analysis and curatorial design to technical deployment—as well as insights from expert participants, we summarize key reflections below.

\subsubsection{Opportunities.}

Our study affirmed the distinctive experiential potential of MR exhibitions, particularly the enhanced embodied sense of immersion when digital content is responsively situated in physical space. This supports findings from earlier research that emphasized MR's capacity to increase visitor engagement through sensory integration and contextualized visual storytelling \cite{Chng_2020_AcceptanceExperienceExpectationsCultural, Jung_2016_AugmentedRealityVisitorExperiences}. While most prior work has focused on in-gallery or museum-based experiences, our deployment in an open urban campus setting demonstrates how these immersive qualities can scale beyond bounded interiors. These observations were reinforced by multiple expert participants, who noted that the seamless blending of virtual and real elements generated a kind of multisensory engagement that traditional media or static MR placements rarely achieve. After experiencing the exhibition, curators and artists emphasized how spatialized interaction and narrative layering contributed to a “felt” sense of presence and aesthetic resonance. In this regard, MR was seen not only as a display technology but as a curatorial material in its own right—capable of shaping perception, movement, and attention. 

\subsubsection{Challenges.} \label{subsec:discussin-challenges}

At the same time, curating such experiences raises significant difficulties. First, site selection in real-world public settings requires careful calibration of spatial affordances, public accessibility, and narrative potential. 

Second, as several experts observed, walking is not merely a navigational act in MR exhibitions—it becomes an active structuring device for narrative continuity and spatial composition. This directly echoes Millard’s concept of \textit{loco-narrative harmony} \cite{Muller_2010_interactivepublicdisplays}, which emphasizes the importance of aligning narrative pacing with spatial progression. Transitions such as turning corners, passing through open space, or arriving at a new zone must be dramaturgically planned to support story immersion and perceptual coherence. This also responds to prior studies that treat visitor trajectory as a central design concept in digital storytelling environments \cite{Benford_2009_coherentjourneysuserexperiences}. 

Third, the interdisciplinary nature of MR curation introduces new demands on curators, designers, and technologists alike. Both our design process and the experts’ prior experiences suggest that curators often need to develop a basic understanding of technical systems, while technologists must adapt their tools and methods to align with curatorial goals and artistic intent. This finding echoes concerns in prior studies about mismatched expectations in co-creative MR projects \cite{Bozzelli_2019_frameworkuser-centricinteractiveexperience,Wei_2025_UsingVirtualRealitySocial}. As one expert (C01) in the online focus group noted, “the discursive power in MR exhibitions tends to shift toward those who control the technology.”

\subsection{MR Curation Strategies to Address the Challenges (RQ2)}


Building on the challenges identified in Section \ref{subsec:discussin-challenges}, this section presents our cross-cutting design strategies that emerged from our MR exhibition design and curation process. These strategies integrate multi-disciplinary insights from expert discussions and articulate how our findings extend prior work.

\subsubsection{Challenge: Complexities of Site Selection in Public Environments}

\paragraph{\textit{\textbullet\ Strategy: Design based on contextual factors.}}

Designing an MR art exhibition in public space, particularly one not originally intended for cultural display, requires more than overlaying content onto pre-existing curatorial infrastructures. Unlike prior MR applications in museums \cite{Gao_2023_VisitingExperienceUsingAR} or heritage sites \cite{Shin_2023_LinkingTrajectoryNarrativeIntent}—where the physical context already carries interpretive significance—we began with a blank spatial canvas: a campus environment and a set of digital artworks. Thus, one of our initial challenges was to identify and select meaningful zones within the campus to host virtual content. 

To address this, we adopted contextualism as a foundational curatorial principle throughout the design process. During the contextual curation phase, we conducted systematic analysis of land-use patterns, circulation flows, and landmarks to understand the functional and structural characteristics of the campus core. This analysis enabled us to assess how existing public-space functions, pedestrian rhythms, and symbolic nodes might support or constrain MR deployment. By comparing these contextual characteristics with our curatorial requirements, we identified areas with the spatial affordances and thematic potential needed to host the exhibition. This approach echoes prior design work \cite{Papageorgopoulou_2021_installationbuildingenhancingcitizens} that highlights the importance of embedding installations within environmental affordances to attract and sustain audience attention as they pass by. In our design,  artworks were intentionally positioned along everyday circulation routes and placed within vibrant public nodes, ensuring that the exhibition integrated naturally into campus life while remaining accessible and legible to visitors.

This approach reflects what contextualist theories across disciplines have emphasized: that meaning is co-constructed by place and experience. In HCI and design research, this aligns with situated interaction theory, which holds that action and interpretation emerge from specific environmental contexts \cite{suchman1987,brown1989situated}. Our strategy applies these principles by treating spatial analysis as a curatorial foundation. As one expert in the focus group noted, \textit{“You didn’t just use space—you need to read it first”} (C01), highlighting the shift from space as neutral backdrop to space as narrative partner.

\paragraph{\textit{\textbullet\ Strategy: Determine MR interfaces based on the physical characteristics of the space.}}


Identifying suitable locations for MR content in public environments requires careful attention to the physical characteristics of each space and its capacity to support different forms of virtual content. Public spaces already contain architectural structures, circulation patterns, and material constraints, meaning that not every area can accommodate every type of MR interface. During the curatorial process, we therefore assessed each site's spatial affordances—such as volume, geometry, visibility, openness, and surface properties—to determine where specific content types could be appropriately integrated spatially. For instance, building façades, as vertical and planar surfaces, naturally support anchored or overlay-based content; indoor spaces with high ceilings can host immersive or overhead-enveloping pieces; and open plazas are better suited for large-scale volumetric 3D works due to unobstructed viewing angles.

Experts observed from prior experience that when MR content is not aligned with a site’s spatial affordances—such as its volume, geometry, visibility, material surfaces, or openness—the resulting experience becomes distracting or cognitively incoherent. Across roles, professionals agreed that MR presentation interfaces must align with the physical characteristics of the hosting space. This consideration also echoes prior research on interface design in MR environments \cite{Han_2023_CreateAmbientMixedRealitya}, which highlights that selecting the appropriate interface is critical for enhancing interaction with real-world objects and supporting perceptual coherence. Studies have shown that aligning virtual content with existing physical structures increases perceptual stability and reduces cognitive overload by avoiding unnecessary competition between digital and physical elements \cite{Li_2023_Object-CenteredUserInterfaceHead-Worn,Han_2023_CreateAmbientMixedRealitya}. These insights underscore that in public-space MR exhibitions, spatial–media alignment is not merely a design preference but a prerequisite for creating stable, legible, and engaging experiences.


\subsubsection{Challenge: Narrative Disruption in Walking-Based MR Experiences}

\paragraph{\textit{\textbullet\ Strategy: Design from the visitor’s perspective.}}

To achieve a cohesive experience across the spatial journey and avoid narrative disruption, we designed the exhibition from the visitor’s point of view—accounting for walking trajectories, sightlines, and spatial transitions. This strategy aligns with Benford et al.’s trajectory framework and with Shin and Woo’s design recommendations \cite{Benford_2009_coherentjourneysuserexperiences,Shin_2023_LinkingTrajectoryNarrativeIntent}, both of which emphasize that MR experiences should be understood from the user’s movement. Our work extends this idea by offering more concrete guidance for outdoor urban environments, where visibility, accessibility, and viewing distance are less controllable than in indoor settings. In our design, artworks were positioned at eye level to support natural engagement, and routes were curated to allow for visual continuity and a gradual thematic reveal. We placed intricate artworks at eye level to support close observation, while pieces that did not require detailed viewing were positioned higher. We also avoided locations where environmental structures, such as hedges or low barriers, would block visitor access or reduce visibility. These choices ensured that visitors could naturally perceive, approach, and engage with each artwork without interruption.

Expert feedback affirmed the value of this visitor-centered arrangement.  At the same time, they pointed out areas where further refinement would be beneficial—particularly the risk of information overload when multiple artworks compete for attention within a shared visual field. They emphasized the importance of highlighting key content through careful sightline management and compositional clarity. This feedback suggests that designing from the visitor’s perspective requires even more fine-grained control of visual emphasis \cite{Gardony_2020_realitypromiseschallenges,Kelley_2024_AspectsEffectiveVisualCueing}, and may benefit from adaptive strategies that accommodate different visitors’ needs and viewing behaviors \cite{Lu_2024_MixedRealityEmpiricalUser}.


\paragraph{\textit{\textbullet\ Strategy: Integrate narrative logic with space.}}

Another strategy involved binding narrative logic closely to the qualities of each physical site. To integrate digital artworks into the public environment, it was essential to identify locations where the artworks could meaningfully "land," ensuring that each piece aligned with how people cognitively understood the space. In the contextual curation process, we first examined the narrative themes of artwork groups and then paired them with areas that conveyed similar inherent meanings, which formed our four thematic zones (see Table \ref{tab:curatorial zones}). For example, the lawn—often associated with life and nature—was selected for artworks related to growth, while the front plaza, a symbolic academic node, was matched with pieces reflecting themes of thought, wisdom, or culture. During the detailed artwork-space matching stage, we focused on the essence of each artwork and placed it in the specific campus location that best reflected its narrative. Unlike many prior MR exhibitions \cite{Shin_2023_LinkingTrajectoryNarrativeIntent,Park_2024_AttractiveSceneryKeyAugmentinga}, our narrative curation did not depend on an explicit storyline but instead emerged through spatial–narrative alignment.

These decisions ensured that spatial ambience and symbolic identity complemented each other to reinforce the narrative interpretation of the artworks. One curator (C02) remarked that the spatial sequence created “\textit{a feeling of being guided into different worlds without being told explicitly what story was entering},”indicating that narrative cohesion arose not from verbal cues but from well-structured spatial progression. At the same time, experts also raised concerns about the limits of this approach in more complex environments. They pointed out that when the surrounding context becomes denser or less predictable, visitors may not follow the intended route unless clearer narrative cues are provided. Without such cues, they cautioned, deviations in visitor pathways could lead to breaks in narrative continuity. This indicates that in denser or less predictable settings, stronger and more explicit alignment between narrative cues and spatial context is needed to sustain a coherent visitor experience.

\subsubsection{Interdisciplinary Demands of MR Curation}

\paragraph{\textit{\textbullet\ Strategy: Establish early and ongoing interdisciplinary collaboration.}}

In developing the MR exhibition, it became evident that no single disciplinary expertise was sufficient to achieve complex spatial narratives and immersive experiences. The project required collaboration across urban design, VR development, HCI research, graphic design, and media art, showing that interdisciplinary integration was not peripheral but foundational to the design process.

Our expert discussion further confirmed this insight. Participants from backgrounds in art, HCI, engineering, and operations consistently emphasized that MR exhibitions rely on sustained collaboration across domains, rather than supplementary contributions added at later stages. This echoes earlier calls for interdisciplinary strategies in digital curation \cite{Wei_2025_UsingVirtualRealitySocial}. 

Such collaboration spans multiple dimensions. Spatial design concerns layout planning and the integration of digital layers with the physical site. Technology involves platform development, device adaptation, and interaction design. Artistic creation includes 3D visual production, soundscapes, and immersive narratives. Operations address exhibition maintenance and venue coordination. In addition, curatorial expertise—such as audience experience design and narrative framing—cuts across these domains, informing decisions at every stage. These dimensions are interdependent and collectively indispensable for MR exhibition practice.

A key guideline emerging from the expert discussion was initiating collaboration early in the design process. Experts stressed that artists and curators should not only engage with, but also develop a working understanding of technical capacities. This enables them to balance artistic intent with technological constraints, creating more immersive experiences while preserving artworks’ expressive intentions despite existing technical barriers.

This principle became particularly evident in our MR exhibition curation and spatial deployment. The cross-disciplinary design team—comprising curators, urban designers, and technologists—established a shared vision for the MR exhibition at an early stage through continuous communication. However, artists were not involved during the exhibition design procedure itself. As a result, when they later visited the exhibition, some expressed that the final presentation exceeded their expectations. In contrast, others noted that the way their works were ultimately displayed diverged entirely from what they had initially imagined. This situation underscores the risks of excluding artists from the early collaborative process and highlights the potential gap between artistic vision and technical realization.

\subsection{Strategies Effectively Foster the General Audience's Experience (RQ3)}

According to feedback from participants in the general audience, the alignment of the digital content with the physical environment was positively highlighted and addressed throughout the visit, with participants describing the experience as \textit{"feels real," "natural,"} and \textit{"coherent."} The reasons for this positive feedback stemmed from two aspects, which reflect the effectiveness of the strategies we applied during the design process.

First, participants expressed that many artworks felt naturally integrated into the environment due to the seamless geometric alignment between the digital content and the space’s characteristics. This was part of our design strategy during the contextual curation process—\textit{Determine Interfaces with Physical Space Characteristics}—where we purposely selected spaces for artworks based on the geometric characteristics of both the space and the artworks.

Second, participants also expressed that the natural feel came from the digital-physical match, which aligned with their everyday expectations. For instance, phrases like "mushrooms on lawns," "animals on grass," and "human-scale Magu right at the street corner" conveyed how the placements made sense to them. This aspect was validated by our design strategy—\textit{Integrating Narrative Logic with Space}—as confirmed through user study results. Whether individual artworks or groups, we ensured that each was considered for its conformity with the logical characteristics of the surrounding environment and the specific exhibition zones.

\subsection{Positioning Contextualism as a Curatorial and Design Principle for MR Exhibitions}

Together, these design strategies illustrate the power of a contextualist approach to MR curation. We now discuss how contextualism, as a theoretical lens, shaped our process and can inform future practice. Contextualism has long been established in architecture, urban design, and curatorial theory as an approach that emphasizes dialogue between artworks and their environments, where space becomes an active component of meaning-making \cite{Elshater_2025_FormationTextualRepresentation}.

Our findings show that contextualism can serve as a guiding principle for MR exhibitions in such environments. In our campus-based MR exhibition, contextualism functioned as a theory–practice framework for embedding digital content within public spaces in ways that resonate with spatial character and audience perception. This principle was validated through both user feedback and expert perspectives. Across the stages of contextual curation, curatorial design, and technical implementation, our project demonstrates how contextual analysis, visitor-centered spatial sequencing, and spatial–media alignment collectively embed MR content within the affordances and meanings of a site. These practices reveal contextualism not as a discrete design step or a purely theoretical construct, but as the foundational and operationalizable design principle that informed placement decisions, narrative sequencing, and digital–physical integration.

As environments grow more complex and visitor movement becomes less predictable, MR designers will need contextual strategies that link spatial affordances, narrative cues, and technical interfaces. Therefore, these insights position contextualism as a practical curatorial and design principle for future MR exhibitions in public spaces.
 
\subsection{Limitations and Future Work}

\subsubsection{Exhibition Limitations.}

Findings from both the focus group and user study reveal several limitations shaped by the technical conditions of MR exhibitions. The most prominent challenges were related to tracking reliability and environmental sensitivity. Experts noted that outdoor MR experiences depend heavily on a stable spatial scan; changes in lighting, weather, or surface conditions could disrupt anchoring, raise safety concerns, and limit the range of movements or viewing angles that are feasible in practice. Additionally, participants from the general audience reflected that the headset display sometimes distorted the physical environment in their perception, causing confusion.

Rendering performance also constrained the exhibition. Limited computational power occasionally introduced visual artifacts—such as grain or simplified textures—that broke immersion and created a noticeable gap between artistic intention and system capability. These constraints further affected authorship: when certain effects could not be rendered, technical teams had to reinterpret or modify artworks, meaning the final presentation sometimes reflected what the system could support rather than what the artist originally envisioned.

Finally, device-dependent accessibility restricted public reach. While head-mounted displays enabled a strong sense of immersion, their cost, limited availability, and occasional discomfort for particular groups of users reduced the number of visitors able to experience the exhibition. Experts highlighted that running an MR exhibition of this scale requires substantial operational resources, making long-term deployment difficult.

\subsubsection{Study Limitations.}

There are several limitations in our study. First, the scope was limited to a single campus-based exhibition, which may not capture the diversity of public spaces. The findings may not generalize directly to more complex environments with denser crowds, higher noise levels, or different architectural characteristics. Future studies should examine MR exhibitions across varied public contexts to test the robustness of our strategies under diverse spatial and social conditions.

Second, the participant sample, while varied, cannot represent all audience groups or professional perspectives. Our study combined insights from experts and general audiences, but broader engagement with diverse demographics, cultural backgrounds, and professional domains is needed. Longitudinal deployments would also provide valuable insights into sustained engagement and the evolving role of MR in public space. Audience familiarity with the site should be considered as an important factor in future evaluations. In location-based MR exhibitions, participants with different levels of attachment and prior experience of the site are likely to interpret and respond to digital artworks in distinct ways. Investigating these differences would offer deeper insight into how MR experiences mediate between personal spatial knowledge, emotional connection, and artistic interpretation.

Taken together, these limitations highlight the need for future work to develop more robust technical pipelines, test MR exhibitions across diverse urban environments, and include broader and longer-term participation. Such work will strengthen the generalization of curatorial strategies and advance the integration of MR into public space.

\section{Conclusion}

This study explored how MR exhibitions can be curated in public spaces through a contextualist approach. By staging a campus-based MR exhibition and integrating expert and audience evaluations, we identified opportunities such as deepening immersion and lowering barriers to art engagement, while also revealing challenges related to site selection, narrative continuity, and interdisciplinary coordination. Our findings highlight contextualism as a curatorial principle that anchors digital content in site-specific qualities, reinforcing meaning-making and fostering deeper engagement with the artworks. Building on this framework, we proposed design strategies to address the challenges encountered during curation, including designing based on contextual factors, determining MR interfaces according to the physical characteristics of space, designing from the visitor’s perspective, and integrating narrative logic with spatial context. Taken together, this work provides practical strategies and theoretical grounding for embedding MR exhibitions into everyday environments. Future research should test these strategies in diverse public contexts, strengthen technical infrastructures, and broaden participation, ensuring that MR’s potential to enrich cultural engagement in public space is fully realized.

\begin{acks}
This work was supported in part by the National Key Research and Development Program of China under Grant No. 2024YFC3307602, and the Guangdong Provincial Talent Program, Grant No. 2023JC10X009.

We thank the study participants for their time and insights.

\end{acks}


\clearpage
\FloatBarrier

\bibliographystyle{ACM-Reference-Format}
\bibliography{Reference}


\clearpage
\appendix
\section{Appendix}

\subsection{Role-Specific Prompts for the Focus Group}
\label{Apex:expert prompts}
\begin{table}[H]
\footnotesize
\caption{The table summarizes the open-ended questions prepared for each expert role before the online session. These prompts guided the semi-structured discussions and ensured coverage of spatial, curatorial, artistic, and technical dimensions relevant to MR exhibition design.}
\label{tab:expert_prompts}
\centering
\renewcommand{\arraystretch}{1.6}

\begin{tabular}{@{}p{0.22\columnwidth} p{0.73\columnwidth}@{}}
\toprule
\textbf{Expert Role} & \textbf{Discussion Prompts (Open-Ended Questions)} \\
\midrule

\textbf{General Reflection} &
1. Which aspects of the exhibition did you find most inspiring or surprising? \par
2. Which elements in the MR exhibition did you find confusing? \par
3. What parts of the MR exhibition should be improved or further explored? \\
\addlinespace[0.4em]

\textbf{Curator} &
1. How do you evaluate MR as an exhibition medium in terms of narrative, spatial layout, and audience guidance? \par
2. What kinds of new curatorial approaches might MR exhibitions need? \par
3. If situated in public spaces (e.g., museums, plazas, parks), what curatorial adjustments would be necessary? \\
\addlinespace[0.4em]

\textbf{Architects \& Urban Designers} &
1. What role can MR play in urban space? \par
2. How can MR contribute to urban narratives or spatial memory? \par
3. How does this MR exhibition inform your thinking about future mixed physical--digital cities? \\
\addlinespace[0.4em]

\textbf{Artists} &
1. How should the fit between content and space be considered in MR artistic creation? \par
2. Compared with traditional media, what unique expressive possibilities does MR offer? \par
3. Do MR artworks require active viewer engagement? \\
\addlinespace[0.4em]

\textbf{Technologists} &
1. Which technological components most strongly shape user experience? \par
2. How does MR expand possibilities for artistic creation and audience experience, and where are its limits? \par
3. In outdoor or complex real-world settings, what are the key technical challenges? \par
4. MR development spans content, devices, and platforms---are current collaboration mechanisms effective and how might they improve? \\
\addlinespace[0.4em]

\textbf{Co-creation Prompts} &
1. How should MR experiences complement physical environments in future public spaces? \par
2. If designing an MR experience for an urban site, which qualities would you prioritize (e.g., participation, narrative, accessibility, management)? \par
3. What collaboration mechanisms among artists, designers, technologists, and curators seem most effective for MR projects? \\
\bottomrule
\end{tabular}

\end{table}

\clearpage
\subsection{Demographic profile of participants from the general audience (N=71)}
\label{Apex:Demographic profile of the participants (N=71)}
\begin{table}[H]
\centering
\caption{Demographic profile of the participants (N=71).}
\label{tab:audience demographics}
\begin{tabular}{lcc}
\toprule
\textbf{Demographic Variables} & \textbf{Frequency} & \textbf{Percent (\%)} \\
\midrule
Sample size & 71 & 100.0 \\
\midrule
\textit{Age} & & \\
\quad 18-25 years & 3 & 4.2 \\
\quad 26-35 years & 31 & 43.7 \\
\quad 36-45 years & 27 & 38.0 \\
\quad 46-55 years & 6 & 8.5 \\
\quad 55+ years & 4 & 5.6 \\
\midrule
\textit{Campus Familiarity} & & \\
\quad Very Familiar & 27 & 38.0 \\
\quad Familiar & 11 & 15.5 \\
\quad Neutral & 15 & 21.1 \\
\quad Unfamiliar & 4 & 5.6 \\
\quad Very Unfamiliar & 14 & 19.7 \\
\midrule
\textit{MR/AR/VR Experience} & & \\
\quad Never Used & 13 & 18.3 \\
\quad Rarely Used & 35 & 49.3 \\
\quad Sometimes Used & 23 & 32.4 \\
\midrule
\textit{Design/Art Background} & & \\
\quad No Background & 23 & 32.4 \\
\quad Some Background & 48 & 67.6 \\
\midrule
\textit{Digital Art Knowledge} & & \\
\quad No Knowledge & 7 & 9.9 \\
\quad Basic Knowledge & 8 & 11.3 \\
\quad Intermediate Knowledge & 31 & 43.7 \\
\quad Advanced Knowledge & 22 & 31.0 \\
\quad Professional Knowledge & 3 & 4.2 \\
\bottomrule
\end{tabular}
\end{table}

\clearpage
\subsection{The List of Interview Participants}
\label{Apex:List of Interview Participants}
\begin{table}[H]
\centering
\footnotesize
\caption{The List of Interview Participants.}
\label{tab: Interview Participants}
\begin{tabular}{@{}l l c c c@{}}
\toprule
\textbf{ID} & \textbf{Gender} & \textbf{Age} & \textbf{Nationality} \\
\midrule
P02 & Male             & 26-35 years   & France \\
P15 & Male             & 26-35 years   & China  \\
P19 & Female           & 18-25 years   & China  \\
P20 & Male             & 26-35 years   & China  \\
P21 & Female           & 26-35 years   & China  \\
P29 & Male             & 18-25 years   & China  \\
P31 & Female           & 26-35 years   & China  \\
P32 & Female           & 26-35 years   & China   \\
P35 & Male             & 26-35 years   & China   \\
P36 & Female           & 26-35 years   & China   \\
P49 & Male             & 18-25 years   & China   \\
P50 & Male             & 26-35 years   & China   \\
P51 & Male             & 26-35 years   & China   \\
P52 & Female           & 26-35 years   & China   \\
P57 & Female           & 18-25 years   & China   \\
P65 & Female           & 26-35 years   & China   \\
P66 & Male             & 26-35 years   & China   \\
P68 & Female           & 26-35 years   & China   \\
P69 & Female           & 36-50 years   & China   \\
P70 & Female           & 26-35 years   & China   \\
P71 & Female           & 26-35 years   & China   \\
\bottomrule
\end{tabular}
\end{table}

\FloatBarrier 
\subsection{The List of Open-end Interview Question}
\label{User_Questionaires}
1.	What are your overall impressions of the exhibition? Which part impressed you the most?

2.	Compared to traditional exhibitions, what do you think is special or innovative about this one?

3.	During your visit, were you still aware of the physical environment? Was the real-world environment enhanced or diminished?

4.	Which parts do you feel most effectively integrated the physical space with virtual content?

5.	Were there any specific design elements (e.g., sound, portals, etc.) that particularly caught your attention?

6.	Were there any parts that confused you or felt inconvenient to experience?

7.	Do you have any suggestions for how we can improve this kind of MR exhibitions?

8. Do you have any other feelings or suggestions you'd like to share?

\end{document}